\documentclass[pra,notitlepage,twocolumn,nofootinbib,superscriptaddress,longbibliography]{revtex4-2}

\usepackage[dvipsnames]{xcolor}
\usepackage{framed}
\definecolor{shadecolor}{rgb}{0.9,0.9,0.9}

\usepackage{mathtools}
\usepackage{amsmath}
\usepackage[shortlabels]{enumitem}

\usepackage{graphicx,epic,eepic,epsfig,amsmath,latexsym,amssymb,verbatim,color}
 
\usepackage{amsfonts}       
\usepackage{nicefrac}       
 
 \usepackage{mathrsfs}

\usepackage{amsmath}
\usepackage{bbm}

\usepackage{tikz}
\usetikzlibrary{chains}
\usetikzlibrary{fit}
\usetikzlibrary{arrows}
\usetikzlibrary{decorations}

\usepackage{epsfig}
\usetikzlibrary{shapes.symbols,patterns} 
\usepackage{pgfplots}

\usepackage[strict]{changepage}
\usepackage{hyperref}
\hypersetup{colorlinks=true,citecolor=blue,linkcolor=blue,filecolor=blue,urlcolor=blue,breaklinks=true}

\usepackage[marginal]{footmisc}
\usepackage{url}
\usepackage{theorem}

\newtheorem{definition}{Definition}
\newtheorem{proposition}{Proposition}

\newtheorem{theorem}[proposition]{Theorem}

\def\squareforqed{\hbox{\rlap{$\sqcap$}$\sqcup$}}
\def\qed{\ifmmode\squareforqed\else{\unskip\nobreak\hfil
\penalty50\hskip1em\null\nobreak\hfil\squareforqed
\parfillskip=0pt\finalhyphendemerits=0\endgraf}\fi}
\def\endenv{\ifmmode\;\else{\unskip\nobreak\hfil
\penalty50\hskip1em\null\nobreak\hfil\;
\parfillskip=0pt\finalhyphendemerits=0\endgraf}\fi}

\newcounter{remark}

\newcounter{example}
\newenvironment{example}[1][]{\refstepcounter{example}\par\medskip\noindent%
\textbf{Example~\theexample #1} }{\medskip}

\mathchardef\ordinarycolon\mathcode`\:
\mathcode`\:=\string"8000
\def\vcentcolon{\mathrel{\mathop\ordinarycolon}}
\begingroup \catcode`\:=\active
  \lowercase{\endgroup
  \let :\vcentcolon
  }

\usepackage{cleveref}
\usepackage{graphicx}
\usepackage{xcolor}

\RequirePackage[framemethod=default]{mdframed}
\newmdenv[skipabove=7pt,
skipbelow=7pt,
backgroundcolor=darkblue!15,
innerleftmargin=5pt,
innerrightmargin=5pt,
innertopmargin=5pt,
leftmargin=0cm,
rightmargin=0cm,
innerbottommargin=5pt,
linewidth=1pt]{tBox}

\newmdenv[skipabove=7pt,
skipbelow=7pt,
backgroundcolor=red!15,
innerleftmargin=5pt,
innerrightmargin=5pt,
innertopmargin=5pt,
leftmargin=0cm,
rightmargin=0cm,
innerbottommargin=5pt,
linewidth=1pt]{rBox}

\newmdenv[skipabove=7pt,
skipbelow=7pt,
backgroundcolor=blue2!25,
innerleftmargin=5pt,
innerrightmargin=5pt,
innertopmargin=5pt,
leftmargin=0cm,
rightmargin=0cm,
innerbottommargin=5pt,
linewidth=1pt]{dBox}
\newmdenv[skipabove=7pt,
skipbelow=7pt,
backgroundcolor=darkkblue!15,
innerleftmargin=5pt,
innerrightmargin=5pt,
innertopmargin=5pt,
leftmargin=0cm,
rightmargin=0cm,
innerbottommargin=5pt,
linewidth=1pt]{sBox}
\definecolor{darkblue}{RGB}{0,76,156}
\definecolor{darkkblue}{RGB}{0,0,153}
\definecolor{blue2}{RGB}{102,178,255}
\definecolor{darkred}{RGB}{195,0,0}

\newcommand{\nc}{\newcommand}
\nc{\rnc}{\renewcommand}
\nc{\lbar}[1]{\overline{#1}}
\nc{\bra}[1]{\langle#1|}
\nc{\ket}[1]{|#1\rangle}

\nc{\ketbra}[2]{|#1\rangle\!\langle#2|}
\nc{\dketbra}[2]{|#1\rangle\!\rangle\!\langle\!\langle#2|}

\nc{\braket}[2]{\langle#1|#2\rangle}

\nc{\kett}[1]{|#1\rangle\!\rangle}
\nc{\braa}[1]{\langle\!\langle#1|}

\nc{\proj}[1]{| #1\rangle\!\langle #1 |}
\nc{\avg}[1]{\langle#1\rangle}
\nc{\rank}{\operatorname{Rank}}
\nc{\smfrac}[2]{\mbox{$\frac{#1}{#2}$}}
\nc{\tr}{\operatorname{Tr}}
\nc{\ox}{\otimes}
\nc{\dg}{\dagger}
\nc{\dn}{\downarrow}
\nc{\cA}{{\cal A}}
\nc{\cB}{{\cal B}}
\nc{\cC}{{\cal C}}
\nc{\cD}{{\cal D}}
\nc{\cE}{{\cal E}}
\nc{\cF}{{\cal F}}
\nc{\cG}{{\cal G}}
\nc{\cH}{{\cal H}}
\nc{\cI}{{\cal I}}
\nc{\cJ}{{\cal J}}
\nc{\cK}{{\cal K}}
\nc{\cL}{{\cal L}}
\nc{\cM}{{\cal M}}
\nc{\cN}{{\cal N}}
\nc{\cO}{{\cal O}}
\nc{\cP}{{\cal P}}
\nc{\cQ}{{\cal Q}}
\nc{\cR}{{\cal R}}
\nc{\cS}{{\cal S}}
\nc{\cT}{{\cal T}}
\nc{\cU}{{\cal U}}
\nc{\cV}{{\cal V}}
\nc{\cX}{{\cal X}}
\nc{\cY}{{\cal Y}}
\nc{\cZ}{{\cal Z}}
\nc{\cW}{{\cal W}}
\nc{\csupp}{{\operatorname{csupp}}}
\nc{\qsupp}{{\operatorname{qsupp}}}
\nc{\var}{{\operatorname{var}}}
\nc{\rar}{\rightarrow}
\nc{\lrar}{\longrightarrow}
\nc{\polylog}{{\operatorname{polylog}}}
\nc{\wt}{{\operatorname{wt}}}
\nc{\av}[1]{{\left\langle {#1} \right\rangle}}
\nc{\supp}{{\operatorname{supp}}}

\nc{\argmin}{{\operatorname{argmin}}}

\def\x{\xi}

\nc{\RR}{{{\mathbb R}}}
\nc{\CC}{{{\mathbb C}}}
\nc{\FF}{{{\mathbb F}}}
\nc{\NN}{{{\mathbb N}}}
\nc{\ZZ}{{{\mathbb Z}}}
\nc{\PP}{{{\mathbb P}}}
\nc{\QQ}{{{\mathbb Q}}}
\nc{\UU}{{{\mathbb U}}}
\nc{\EE}{{{\mathbb E}}}
\nc{\id}{{\operatorname{id}}}

\nc{\CHSH}{{\operatorname{CHSH}}}

\nc{\<}{\langle}
\rnc{\>}{\rangle}

\nc{\rU}{\mbox{U}}

\nc{\ob}[1]{#1}
\nc{\QS}{{\bm{\cS}}}

\nc{\SEP}{{\text{\rm SEP}}}
\nc{\NS}{{\text{\rm NS}}}
\nc{\LOCC}{{\text{\rm LOCC}}}
\nc{\PPT}{{\text{\rm PPT}}}
\nc{\EXT}{{\text{\rm EXT}}}
\nc{\STAB}{{\text{\rm STAB}}}

\nc{\Sym}{{\operatorname{Sym}}}

\nc{\ERLO}{{E_{\text{r,LO}}}}
\nc{\ERLOCC}{{E_{\text{r,LOCC}}}}
\nc{\ERPPT}{{E_{\text{r,PPT}}}}
\nc{\ERLOCCinfty}{{E^{\infty}_{\text{r,LOCC}}}}
\nc{\Aram}{{\operatorname{\sf A}}}

\newcommand{\CPWP}{\text{\rm CPWP}}
\newcommand{\CSPO}{\text{\rm CSPO}}

\usepackage{tikz}
\usepackage{hyperref}
\hypersetup{colorlinks=true,citecolor=blue,linkcolor=blue,filecolor=blue,urlcolor=blue,breaklinks=true}

\makeatletter
\def\grd@save@target#1{%
  \def\grd@target{#1}}
\def\grd@save@start#1{%
  \def\grd@start{#1}}
\tikzset{
  grid with coordinates/.style={
    to path={%
      \pgfextra{%
        \edef\grd@@target{(\tikztotarget)}%
        \tikz@scan@one@point\grd@save@target\grd@@target\relax
        \edef\grd@@start{(\tikztostart)}%
        \tikz@scan@one@point\grd@save@start\grd@@start\relax
        \draw[minor help lines,magenta] (\tikztostart) grid (\tikztotarget);
        \draw[major help lines] (\tikztostart) grid (\tikztotarget);
        \grd@start
        \pgfmathsetmacro{\grd@xa}{\the\pgf@x/1cm}
        \pgfmathsetmacro{\grd@ya}{\the\pgf@y/1cm}
        \grd@target
        \pgfmathsetmacro{\grd@xb}{\the\pgf@x/1cm}
        \pgfmathsetmacro{\grd@yb}{\the\pgf@y/1cm}
        \pgfmathsetmacro{\grd@xc}{\grd@xa + \pgfkeysvalueof{/tikz/grid with coordinates/major step}}
        \pgfmathsetmacro{\grd@yc}{\grd@ya + \pgfkeysvalueof{/tikz/grid with coordinates/major step}}
        \foreach \x in {\grd@xa,\grd@xc,...,\grd@xb}
        \node[anchor=north] at (\x,\grd@ya) {\pgfmathprintnumber{\x}};
        \foreach \y in {\grd@ya,\grd@yc,...,\grd@yb}
        \node[anchor=east] at (\grd@xa,\y) {\pgfmathprintnumber{\y}};
      }
    }
  },
  minor help lines/.style={
    help lines,
    step=\pgfkeysvalueof{/tikz/grid with coordinates/minor step}
  },
  major help lines/.style={
    help lines,
    line width=\pgfkeysvalueof{/tikz/grid with coordinates/major line width},
    step=\pgfkeysvalueof{/tikz/grid with coordinates/major step}
  },
  grid with coordinates/.cd,
  minor step/.initial=.2,
  major step/.initial=1,
  major line width/.initial=2pt,
}
\makeatother

\usepackage{thmtools}
\usepackage{thm-restate}
\usepackage{etoolbox}
\makeatletter
\def\problem@s{}
\newcounter{problems@cnt}

\newcommand{\allproblems}{\problem@s}
\makeatother

\usepackage{amsmath,amsfonts,amssymb,array,graphicx,mathtools,multirow,bm,times,tcolorbox,relsize,booktabs,dsfont}
\usepackage[utf8]{inputenc}
\usepackage[T1]{fontenc}
\usepackage{qcircuit}
\usepackage{soul}
\usepackage{titletoc}
\titlecontents{section}[0pt]{\vspace{5pt}\bfseries}{}{\bfseries}{\hfill\contentspage}[\vspace{5pt}]
\titlecontents{subsection}[15pt]{\normalfont}{}{\normalfont}{\hfill\contentspage}[\vspace{5pt}]

\usepackage[ruled, vlined, linesnumbered]{algorithm2e}

\definecolor{colortwo}{rgb}{0.4,0.77,0.17}
\definecolor{colorthree}{rgb}{0.01,0.51,0.93}

\allowdisplaybreaks

\begin{document}
\title{Enhancement of non-Stabilizerness within Indefinite Causal Order}
\author{Yin Mo}
\author{Chengkai Zhu}
\affiliation{Thrust of Artificial Intelligence, Information Hub,\\
The Hong Kong University of Science and Technology (Guangzhou), Guangzhou 511453, China}
\author{Zhiping Liu}
\affiliation{Thrust of Artificial Intelligence, Information Hub,\\
The Hong Kong University of Science and Technology (Guangzhou), Guangzhou 511453, China}
\affiliation{National Laboratory of Solid State Microstructures, School of Physics and Collaborative Innovation Center of Advanced Microstructures, Nanjing University, Nanjing 210093, China}
\author{Mingrui Jing}
\author{Xin Wang}
\email{felixxinwang@hkust-gz.edu.cn}
\affiliation{Thrust of Artificial Intelligence, Information Hub,\\
The Hong Kong University of Science and Technology (Guangzhou), Guangzhou 511453, China}

\begin{abstract}
In quantum computing, the non-stabilizerness of quantum operations is crucial for understanding and quantifying quantum speed-ups. In this study, we explore the phenomena of non-stabilizerness of the Quantum SWITCH, a novel structure that allows quantum states to pass through operations in a superposition of different orders, outperforming traditional circuits in numerous tasks. To assess its non-stabilizerness, we propose the magic resource capacity of a quantum process to quantitatively examine the non-stabilizerness of general quantum transformations. We find that the completely stabilizer-preserving operations, which cannot generate magic states under standard conditions, can be transformed to do so when processed by the Quantum SWITCH. Furthermore, when considering the impact of noise, although the non-stabilizerness of each path may be annihilated, their superposition could still preserve the overall non-stabilizerness. These findings reveal the unique properties of the Quantum SWITCH and open avenues in research on non-stabilizer resources of general quantum architecture.
\end{abstract}

\date{\today}
\maketitle

\section{Introduction}

Quantum computing utilizes quantum effects such as entanglement and superposition to achieve quantum advantages, enabling it to perform tasks that are challenging for classical computers~\cite{shor1997faulttolerant,renner2022computational,Wu2021strong,knill2005quantum}. For instance, Shor's Algorithm~\cite{Shor1999polynomial} can efficiently factor large integers, a capability that could undermine RSA encryption. Recently, advancements in quantum hardware have demonstrated quantum supremacy~\cite{Arute2019} by efficiently completing computational tasks that would take classical computers prohibitively long times to solve~\cite{Zhong2020quantum}.

\begin{figure}[t]
    \centering
    \includegraphics[width=0.9\linewidth]{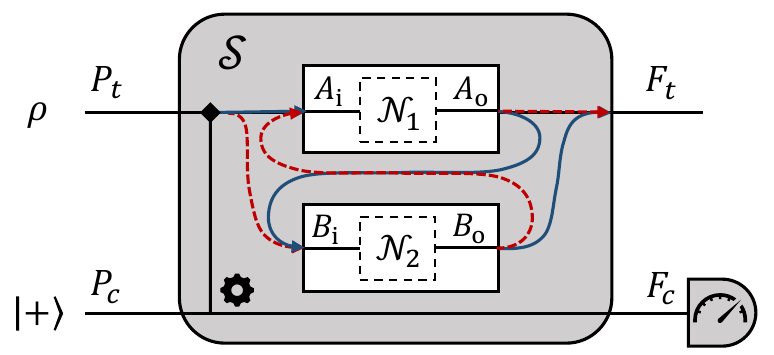}
    \caption{A general illustration of Q-SWITCH using a quantum circuit diagram taking two quantum channels, $\cN_{1,2}$ as inputs, which then acts as a new channel $\QS(\cN_1, \cN_2): P_t \mapsto F_t$. We use the convention that the inward lines into the $\QS$ represent the input spaces, and the outward lines from $\QS$ represent the output spaces. The labels $P_t (F_t)$ and $P_c (F_c)$ indicate the target and the control past (future) systems, respectively, where a $\ket{+}$ state is usually initialized to the control qubit.}
    \label{fig:qswitch}
\end{figure}

Different from traditional quantum computing, where only quantum states are considered coherent and processed through quantum channels in a fixed order, indefinite causal network~\cite{Chiribella_2013,Oreshkov_2012}, which allows the order of quantum states passing through events to be in superposition, has gained widespread attention. A prime example of this is the Quantum SWITCH (Q-SWITCH)~\cite{Chiribella_2013}, a two-slot quantum supermap that uses a control qubit to place input states in a superposition of two processing sequences: first passing through channel $\cN_1$ and then channel $\cN_2$, and vice versa (see Fig.~\ref{fig:qswitch}). This paradigm has been shown to offer significant advantages compared with definite causal networks in various domains, including communication complexity~\cite{Gu_rin_2016, ebler2018enhanced}, channel discrimination~\cite{Bavaresco2021}, quantum thermodynamics~\cite{Felce2020}, quantum computation~\cite{Quintino_2019, renner2022computational}, and quantum metrology~\cite{Zhao2020a}.

It is intriguing to consider whether this novel quantum effect also challenges classical simulation, potentially revealing new quantum advantages. The classical simulability of quantum processes is a critical question, as highlighted by the Gottesman-Knill Theorem~\cite{gottesman2009introduction}, which states that quantum circuits comprising only Clifford gates can be efficiently simulated classically. 
However, the inclusion of non-stabilizer operations~\cite{Gottesman_1999, Zhou_2000}, such as the $T$ gate, makes the complexity of its classical simulation scale exponentially with the number of these operations~\cite{Aaronson_2004}.
To identify such simulation complexity and fully exploit the power of non-stabilizer resources quantitatively in Fault-Tolerant Quantum Computing (FTQC), a resource-theoretic approach known as the quantum resource theory (QRT) of magic states and channels is proposed~\cite{Bravyi2016,Bravyi_2019,Howard_2017,RMP_Chitambar_2019}.

From the perspective of resource theory in FTQC, the capability of generating magic resources is a natural and operational way to quantify the non-stabilizerness of quantum supermaps. To better study this property, we propose `magic resource capacity' for $m$-slot general quantum processes to characterize its non-stabilizerness, which quantifies the ability it could generate quantum channels with non-stabilizerness when taking $m$ free operations as inputs. Based on this, we have analyzed the Q-SWITCH specifically and unveiled some intriguing phenomena.

Conventionally, repeatedly applying a completely stabilizer-preserving operation (CSPO) to a stabilizer state fails to transform it into a magic state. Intriguingly, however, when a stabilizer state is placed in a superposition of passing through two instances of this channel in different orders, the output state can indeed become a magic state. 
We observed that this phenomenon is common by randomly sampling CSPO channels and giving them into the Q-SWITCH. We also introduced an efficiently computable lower bound for its magic resource capacity. 
Besides these, we further explore the impact of the depolarizing noise on the non-stabilizerness of Q-SWITCH. In traditional circuits, accumulated noise in each component could degrade the entire process to one that can be simulated classically. In the Q-SWITCH framework, however, even if depolarizing noise prevents each individual path from maintaining non-stabilizerness, the superposition of these paths ensures that the output state could remain a magic state.

Through these examples, we demonstrate methods based on resource theory to quantitatively analyze the properties of indefinite causal networks. These methods better showcase the quantum utility brought by the superposition of paths in different orders compared to classical computing. As recent experimental progress with photonic and quantum thermodynamic methods has verified and demonstrated the viability of the Q-SWITCH~\cite{Rubino2017,Yin2023experimental,Guo2020experimental,Nie2022,Goswami2018}, our study enhanced our understanding of its potential to leverage unique quantum computational benefits. Furthermore, our findings suggest that indefinite causal networks represent a special resource within quantum resource theory. Novel results may be obtained when considering a more general quantum structure compared to traditional sequential networks.

\section{Preliminaries}
To elucidate the impact of the Q-SWITCH within the QRT of magic states, we begin by recalling the free states and operations within this theory. We also introduce their quantitative measures, which will be instrumental in demonstrating the effects of the Q-SWITCH. 

Consider $\cH_A$ as a finite-dimensional Hilbert space affiliated with a quantum system $A$. We denote $\mathscr{L}(\cH_A)$ as the space of linear operators that map $\cH_A$ onto itself, and $\mathscr{D}(\cH_A)$ represents the set of density operators on $\cH_A$. We call a linear map from $\mathscr{L}(\cH_A)$ to $\mathscr{L}(\cH_B)$ CPTP if it is completely positive and trace-preserving, also known as the general definition of a quantum channel $\cN_{A\to B}$. For a quantum channel $\cN_{A\to B}$, its Choi-Jamiołkowski state is given by $J_{\cN} \equiv \frac{1}{d_A}\sum_{i, j=0}^{d_A-1}\ketbra{i}{j} \ox \cN_{A \to B}(\ketbra{i}{j})$, where $\{|i\rangle\}_{i=0}^{d_A-1}$ is an orthonormal basis in $\cH_A$. 

The QRT of magic states~\cite{Veitch2014,Howard_2017,Seddon_2019,Heinrich_2019,Wang2018,Saxena2022,Beverland2019,Jiang2021,Seddon2021,WWS19,Haug2023,Leone2022,Oliviero2022,Liu2022d,Bu2023a,Haug2023b,Lami2023} establishes a powerful framework to better understand magic states manipulation and quantum speedups. For multi-qubit systems, the free states are the stabilizer states. We denote $\STAB_n$ as the set of $n$-qubit stabilizer states. The free operations are the stabilizer operations (SO) that possess a fault-tolerant implementation in the context of FTQC.
A detailed discussion about the stabilizer formalism can be found in Appendix~\ref{appendix:preliminaries}. 
Notably, a magic monotone called robustness of magic states (RoM) for multi-qubit systems was introduced~\cite{Howard_2017,Seddon_2019,Heinrich_2019}.
The RoM of a $n$-qubit state $\rho$ is defined as
\begin{equation*}\label{def: ROM_states}
    \cR(\rho) = \min_{\Vec{q}}\Big\{ \| \Vec{q} \|_1 \ : \rho=\sum_i q_i \ketbra{s_i}{s_i} \ , \ketbra{s_i}{s_i} \in \STAB_n \Big\},
\end{equation*}
where $\| \Vec{q} \|_1 = \sum_i |q_i|$ denotes the minimal $\ell_1$-norm, and the minimization ranges over possible decompositions of all pure stabilizer states. 
This quantity can be calculated by the linear programming (LP) scheme developed by Seddon and Campbell~\cite{Seddon_2019}. 
We note that $\cR(\cdot)$ is faithful, i.e., $\cR(\rho) \geq 1$ and $\cR(\rho) = 1$ if and only if $\rho\in \STAB_n$.
The free operations were further generalized into the stabilizer-preserving operations (SPO)~\cite{Ahmadi_2018} and the completely stabilizer-preserving operations (CSPO)~\cite{Seddon_2019}. An $n$-qubit quantum channel $\cN_{A\rightarrow B}$ is called CSPO if for any $n$-qubit reference system $R$, 
\begin{equation}
    \forall \rho_{RA}\in\STAB_{2n}, (\id_n \ox \cN_{A\rightarrow B}) (\rho_{RA}) \in \STAB_{2n}.
\end{equation}
The statement is proven to be equivalent to the condition that the Choi operator of $\cN$ satisfies $J_{\cN}\in\STAB_{2n}$.

\section{Generating non-stabilizerness via Q-SWITCH}

According to the definition of CSPO, a stabilizer state cannot be transformed into a magic state by traversing a CSPO channel $\cN$, irrespective of the number of times it is applied. 
Taking into account the possibility of resource generation of quantum switch~\cite{ebler2018enhanced}, it is intriguing to ask if the aforementioned dynamic will be changed.
Specifically, a natural question is when a stabilizer state undergoes a superposition of traversing two copies of this channel in varying orders, essentially passing through Q-SWITCH placement of $\cN$ in Fig.~\ref{fig:qswitch}, if the output state will become a magic state?

\begin{figure}[t]
    \centering
    \includegraphics[width=0.9\linewidth]{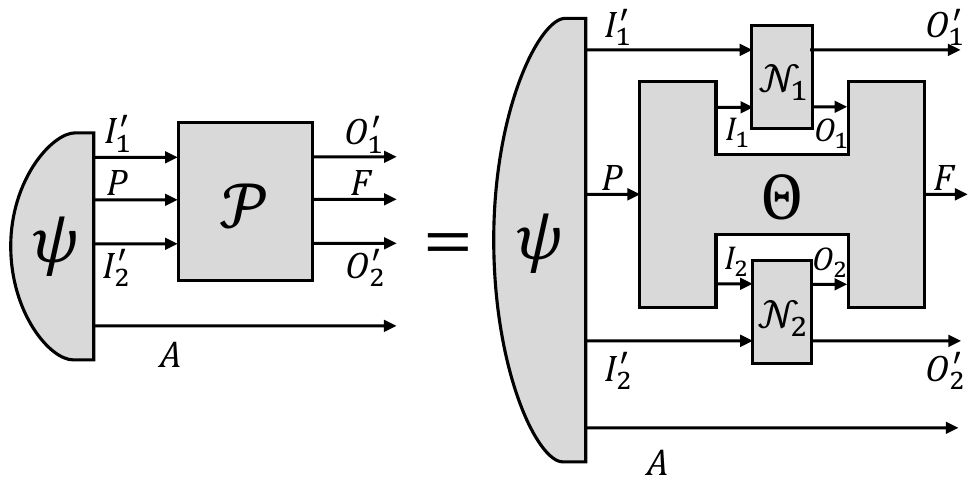}
    \caption{The illustration of a two-slot quantum process $\Theta$, as a transformation mapping two quantum channels $\cN_1, \cN_2$ to a channel $\cP_{I'P\rightarrow O'F}$ where $\cH_{I'} = \cH_{I'_1}\ox\cH_{I'_2}$ and $\cH_{O'} = \cH_{O'_1}\ox\cH_{O'_2}$. The magic resource capacity of $\Theta$ is given by optimizing over all possible CSPO channels $\cN_1, \cN_2$ and stabilizer state $\psi_{A\cI'P}$.}
    \label{fig:n_slot_process}
\end{figure}

Resource-generating capacity is fundamental and operational in various resource theories~\cite{Leifer_2003, bennett2003capacities,Stahlke_2014,Saxena2022}. To address this question and investigate the non-stabilizerness of Q-SWITCH, we propose `magic resource capacity' for $m$-slot general quantum processes to quantify its non-stabilizerness, with the detailed definition given as follows.
\begin{definition}[Magic resource capacity of quantum process]
Given an $m$-slot quantum process $\Theta$ with input channels $\cN_1, \cN_2,...,\cN_m$, the processed channel is written as $\cP_{\cI'P\rightarrow\cO'F}=\Theta(\cN_1, \cN_2,...,\cN_m)$ where each quantum channel $\cN_i$ is from system $I_iI'_i$ to $O_iO'_i$, $\cI'$ is the composite of all $I'_i$ and $\cO'$ is the composite of all $O'_i$. The magic resource capacity of $\Theta$ is defined as
\begin{equation*}
\begin{aligned}
    \widetilde{\cC}(\Theta):= 
    \log\max \Big\{\cR\big((&\mathds{1}_{A} \ox \cP_{\cI'P\rightarrow\cO'F})(\psi_{A\cI' P})\big) :\\
    &\psi_{A\cI' P}\in \STAB,\, \cN_i \in \CSPO,\forall i\Big\}.
\end{aligned}
\end{equation*}
\end{definition}
The logarithms are taken in the base two throughout this paper. This quantity characterizes the maximum magic resource an $m$-slot quantum process could generate by inserting CSPO channels and inputting a stabilizer state into the processed channel. It encompasses all quantum combs with definite causal orders as well as more general processes with indefinite causal orders, e.g., Q-SWTICH. An example of the two-slot quantum process is shown in Fig.~\ref{fig:n_slot_process}, where its magic resource capacity is given by optimizing over all possible CSPO channels $\cN_1$, $\cN_2$, and stabilizer state $\psi_{A\cI' P}$.

Another interesting quantity to consider is when all inserted channels are identical. In this case, we refer to it as the `magic resource capacity with identical inputs' of an $m$-slot quantum process, which we denoted as $\widetilde{\cC}_{\rm ide}(\Theta)$. This measure is particularly special for indefinite causal networks, as the superposition of states passing through identical channels in different orders demonstrates unique quantum phenomena in tasks such as communication~\cite{ebler2018enhanced}. In contrast, in definite causal networks, rearranging the order of identical channels does not impact the outcome. For the Q-SWITCH, we do analyses on both of these quantities.

Suppose two $n$-qubit quantum channels $\cN_1$ and $\cN_2$ have Kraus representations $\{E_i\}_i$ and $\{F_j\}_j$, respectively. ~\citet{ebler2018enhanced} originally gives an effective quantum switched channel $\cS_{\cN_1,\cN_2}:=\QS_n(\cN_1, \cN_2)$ having a Kraus representation $\{W_{ij}\}_{ij}$,
\begin{align}\label{eq: Wij_N1}
    W_{ij} = \ket{0}\bra{0}_c\otimes E_i^{(2)}F_j^{(1)} + \ket{1}\bra{1}_c\otimes F_j^{(1)}E_i^{(2)},
\end{align}
where $c$ is the control system within the Q-SWITCH. To simplify our study on its magic resource capacity, we calculate the following quantity as a lower bound on $\widetilde{\cC}(\QS)$.
\begin{equation*}
\begin{aligned}
    \widetilde{\cC}'\big(\QS_n\big)& := \log\max_{\cN_1,\cN_2} \Big\{\cR\big(\QS_n^+(\cN_1, \cN_2)\ox \mathds{1}_n)(\psi_{2n})\big) + \\
    &\cR\big(\QS_n^-(\cN_1, \cN_2)\ox \mathds{1}_n)(\psi_{2n})\big) \,:\, \psi_{2n}\in \STAB_{2n} \Big\},
\end{aligned}
\end{equation*}
where $(\QS_n^+(\cN_1, \cN_2)\ox \mathds{1}_n)(\psi_{2n})$ is the unnormalized post-measurement state when the control state is measured in Fourier basis and the result is `$+$', so is $\QS_n^{-}(\cdot)$. The maximization ranges over all quantum channels in CSPO, i.e., $\cN_1,\cN_2\in\CSPO$. This quantity characterizes the average maximum magic resource generated by inserting two quantum channels in the Q-SWITCH and inputting a stabilizer state. 
The detailed discussion can be found in Appendix~\ref{appendix:magic_IDC}.

Based on this, we obtain our core result, which demonstrates that the Q-SWITCH is a non-stabilizer resource.
\begin{theorem}\label{thm:main_thm}
    The magic resource capacity of Q-SWITCH is non-zero.
\end{theorem}

We prove the theorem by constructing an efficiently computable lower bound on the magic resource capacity of Q-SWITCH. When $\cN_1$ is a Hadamard gate and $\cN_2$ a channel with Kraus operators $\big\{K_0 = 1/2(-i\ketbra{0}{0}+\ketbra{0}{1}-i\ketbra{1}{0}+ \ketbra{1}{1}), K_1 = 1/\sqrt{2}(-i\ketbra{0}{0}-\ketbra{0}{1})\big\}$, we get that for qubit-to-qubit channels, $\widetilde{\cC}\big(\QS_1\big) \geq 1.138$. We point out that the optimization in $\widetilde{\cC}'(\QS_n)$ is still difficult; to address this, we propose an algorithm based on the standard see-saw procedure. The algorithm consists of three semidefinite programs (SDPs) and is detailed in Appendix~\ref{appendix:magic_IDC}.

\begin{figure}[t]
    \centering
    \includegraphics[width=.95\linewidth]{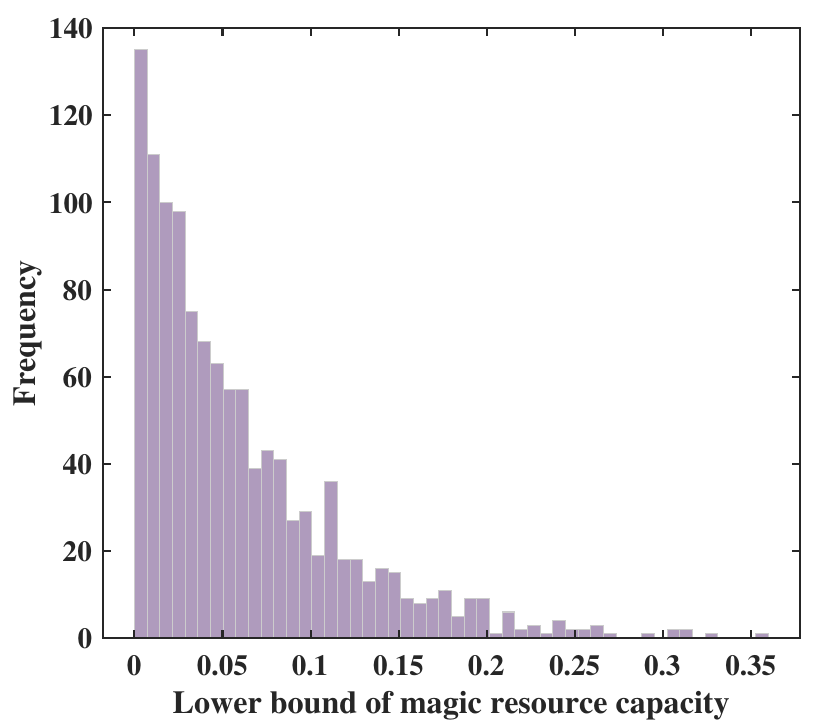}
    \caption{The frequency distribution of lower bound of the magic resource capacity of switched CSPO channel $\QS(\cN_i,\cN_i)$, where $ \cN_i \in \{\cN_i\}_i$ is a set of sampled CSPOs. Here we only count results whose lower bounds of magic resource capacity are larger than $0$.}
    \label{fig:sample_cspo}
\end{figure}

For the `magic resource capacity with identical inputs', the calculation becomes more complicated as the see-saw method cannot be used. 
In this case, we find that when the inserted channel is $\cN_2$ as given above, we could get $\widetilde{\cC}_{\rm ide}\big(\QS_1\big) \geq 0.632$. Besides, we examine the non-stabilizerness generating power of Q-SWITCH by firstly performing a sampling test on $2300$ qubit-to-qubit CSPO channels $\{ \cN_i \}_i$, while implementing Q-SWITCH on two copies of each $\cN_i$ to obtain $\cS(\cN_i, \cN_i)$. Here we use the channel robustness of non-stabilizerness for an $n$-qubit quantum channel $\cN_{A\rightarrow B}$ which is defined as
\begin{equation*}\label{def:channel_r}
    \cR_*(\cN) := \min_{\cN_{\pm} \in \text{CSPO}} \big\{ 2p+1: \cN=(1+p)\cN_+ - p \cN_-, p\geq 0\big\}.
\end{equation*}
Note that the channel robustness of non-stabilizerness can be calculated by a linear programming (LP) scheme developed by Seddon and Campbell~\cite{Seddon_2019}.

In detail, the CSPOs are obtained by sampling $100000$ random qubit-qubit channels by the proposed measures in~\cite{Kukulski_2021}, e.g., generating random Choi operators, and retaining those whose channel robustness is $1$. The numerical calculations are implemented in MATLAB~\cite{MATLAB} with the interpreters CVX~\cite{cvx} and QETLAB~\cite{qetlab}. We compute the lower bound of the magic resource capacity of the switched channel $\QS(\cN_i,\cN_i)$ for each sampled CSPO based on the SDP in Eq.~\eqref{Eq: Magic_CPTN}. There are $1170$ CSPOs generating non-stabilizerness after employing Q-SWITCH, where the maximal value of the lower bound of magic resource capacity among them is $0.357$. We present the frequency distribution of the lower bound of the magic resource capacity of these CSPOs in Fig.~\ref{fig:sample_cspo}. This result shows that Q-SWITCH enables almost half of the sampled CSPOs to generate magic resources, which identifies that Q-SWITCH can generally generate non-stabilizerness resources from stabilizerness.

To better understand this phenomenon, we provide a specific example and calculate its output under the Q-SWITCH, showing that by employing Q-SWITCH, we obtain a magic operation from the original free operations. For the qutrit example, we utilize the discrete Wigner functions~\cite{WOOTTERS19871,Gross_2006a,Gross_2006b}; the details are provided in Appendix~\ref{appendix:qutrit_case}.

\begin{example}
Consider a qubit-to-qubit quantum channel $\cN_{p}$ with its Kraus operators given as $K_0 = \sqrt{\frac{p}{2}}(\ketbra{0}{0}+\ketbra{0}{1})$, $K_1 = \sqrt{\frac{p}{2}}(\ketbra{1}{1}-\ketbra{1}{0})$, $K_2 = \sqrt{1-p}TH$, where $H,T:={\rm diag}(1, e^{i\pi/4})$ are qubit Hadamard gate and $T$ gate respectively.
When $p\in[0.29, 0.59]$, although $\cN_{p}$ is a CSPO, a magic state could be generated from $\QS(\cN_{p}, \cN_{p})$ acting on a stabilizer state.
\end{example}

\begin{figure}[t]
    \centering
    \includegraphics[width=0.95\linewidth]{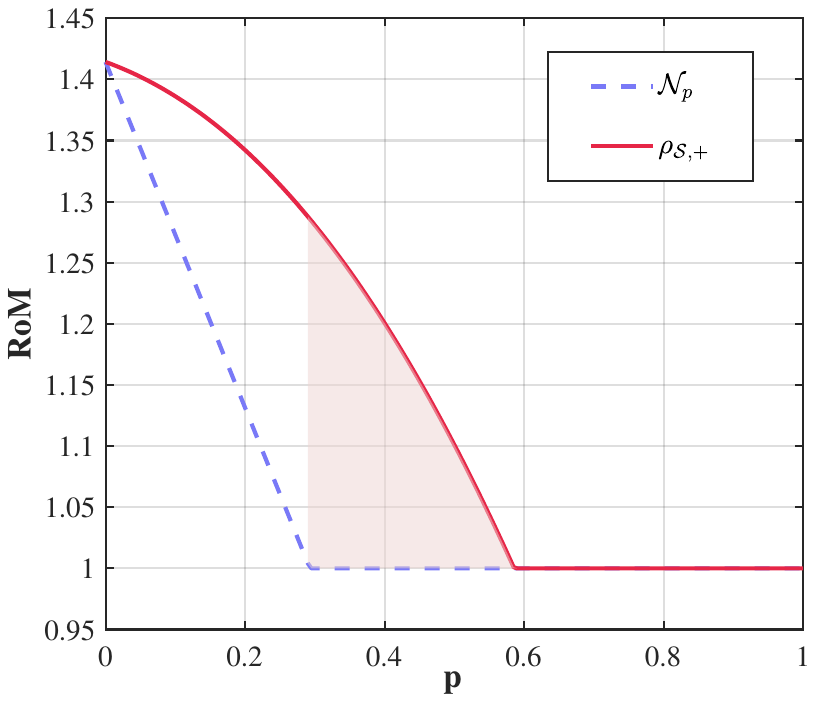}
    \caption{An example for Theorem~\ref{thm:main_thm} in the qubit scenario. The channel robustness of $\cN_p$ is depicted in the blue line, and the RoMs of $\rho_{\QS,+}(p)$ is depicted in the red line, with $p$ ranging in $[0,1]$. The pink area represents the cases where a magic state could be generated from $\QS(\cN_{p}, \cN_{p})$ with a CSPO $\cN_p$ and a stabilizer input state.}
    \label{fig:qubit_qs_seq}
\end{figure}

As shown by the blue line in Fig.~\ref{fig:qubit_qs_seq}, we calculate the robustness of channel $\cN_{p}$. 
It can be seen that when $p\gtrsim 0.29$, $\cN_{p}$ becomes a CSPO.
Then we show that one can generate a magic state by inputting a stabilizer state into $\QS(\cN_{p}, \cN_{p})$. We use $\ket{+}$ as the target input state, set the control state to be $\ket{+}$ and measure it in the Fourier basis $\{\ket{+}, \ket{-}\}$ at last.
Denoting $\ket{\alpha_{ij}} = K_i K_j\ket{+}$, the output state of the whole process can be calculated as

\begin{align*}
    \nonumber\QS(\cN_{p}, \cN_{p})\Big(&\ket{+}\bra{+}\otimes \ket{+}\bra{+}_c \Big) = \ket{+}\bra{+}_{c}\otimes\Big(\sum_i\ket{\alpha_{ii}}\bra{\alpha_{ii}}\Big)\\
    \nonumber&+\frac{1}{2}\Big(\ket{0}\bra{0}_{c} + \ket{1}\bra{1}_{c}\Big) \ox \Big(\sum_{i \neq j}\ket{\alpha_{ij}}\bra{\alpha_{ij}}\Big)\\
    &+\frac{1}{2}\Big(\ket{0}\bra{1}_{c} + \ket{1}\bra{0}_{c}\Big) \ox \Big(\sum_{i \neq j}\ket{\alpha_{ij}}\bra{\alpha_{ji}}\Big).
\end{align*}
At this point, when we measure the control state and obtain the `$+$' result, the conditional final state (unnormalized) is
\begin{align*}
    \rho_{\QS,+}(p) = \, \sum_i\ket{\alpha_{ii}}\bra{\alpha_{ii}}
    + \frac{1}{2}\sum_{i \neq j}\Big(\ket{\alpha_{ij}}\bra{\alpha_{ij}} + \ket{\alpha_{ij}}\bra{\alpha_{ji}} \Big).
\end{align*}

Using a similar technique, we can compute $\rho_{\QS,-}(p)$. 
The RoMs of $\rho_{\QS,+}(p)$ and $ \rho_{\QS,-}(p)$ could then be calculated by solving the LP~\cite{Seddon_2019, Howard_2017} after inserting the formulae of $K_0$, $K_1$, and $K_2$.
In the qubit scenario, we find $\cR(\rho_{\QS,-}(p)) = 1$ always holds, which means $\nonumber\rho_{\QS,-}(p)$ is always a stabilizer state independent of $p$. Therefore, we only show the $\cR(\rho_{\QS,+}(p))$ in this figure. 
In Fig.~\ref{fig:qubit_qs_seq}, the red line represents $\cR(\rho_{\QS,+}(p))$ and the blue line represents $\cR_*(\cN_p)$, both of which vary with the parameter $p\in[0,1]$. The light pink shadow area shows that when $p$ is in the range $[0.29, 0.59]$, one can obtain a magic state with the measurement outcome `$+$', even though the preparation of the state $\ket{+}$, the channel $\cN_{p}$ and the measurement on the control system are all stabilizer operations.

\section{Maintaining non-stabilizerness under noise}\label{main-2}
Quantum advantage is significantly affected by noise; high noise levels will erase quantum information, eliminating the effects that are hard to simulate classically.
In the case of the Q-SWITCH, it is interesting to ask when the noise makes the process on each path degraded to CSPO, could the superposition of two paths still result in a non-CSPO process? In this work, we investigate how the $T$ gate is affected by a common quantum noise model, the depolarizing channel, shown as $\cD_p(\rho) = p\frac{I}{d} + (1-p)\rho$.

Specifically, we compare the following three scenarios:
\begin{enumerate}
    \item[1)] A $T$ gate sequentially passes through two depolarizing channels denoted as $\cT^{\prime}_p = \cD_p \circ \cD_p \circ T$.
    \item[2)] A $T$ gate is in a superposition of traversing through two $p$-depolarizing channels in different orders, denoted as $\QS(\cD_p, \cD_p) \circ T$.
    \item[3)] Q-SWITCH applied to two $\sqrt{T}$ gates each followed by a $p$-depolarizing channel, denoted as $\QS(\cD_p \circ \sqrt{T}, \cD_p \circ \sqrt{T})$.
\end{enumerate}

\begin{figure}[t]
    \centering
    \includegraphics[width=1\linewidth]{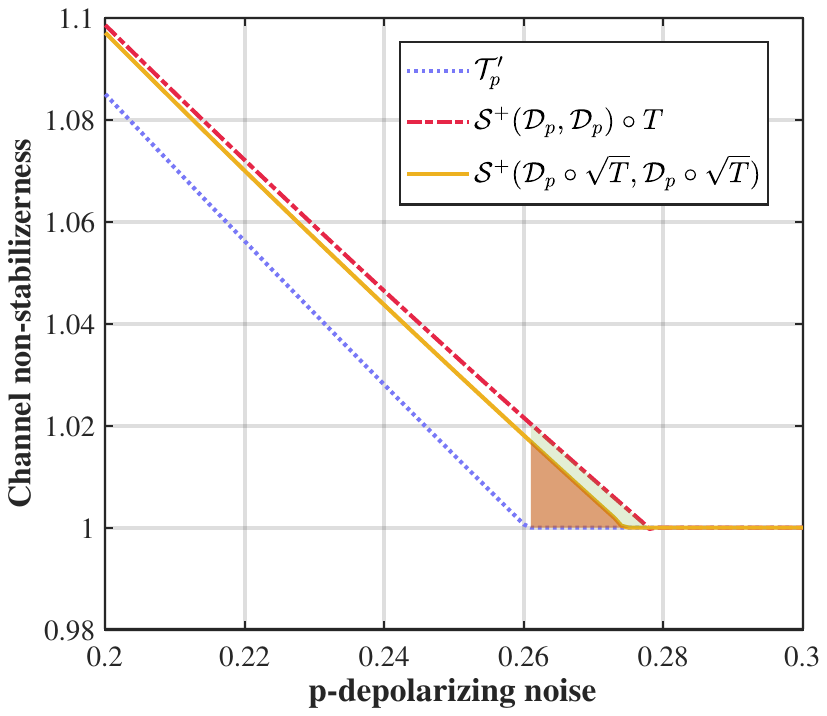}
    \caption{Comparison between the non-stabilizerness of $T$ gate after the depolarizing noise under Q-SWITCH, the non-stabilizerness of $\cD_p \circ \sqrt{T}$ under Q-SWITCH and the non-stabilizerness of $T$ gate after two sequential depolarizing noises. This figure depicts the channel robustness of $\QS^{+}(\cD_p, \cD_p) \circ T$, $\QS^{+}(\cD_p \circ \sqrt{T}, \cD_p \circ \sqrt{T})$ and $\cT^{\prime}_p$ in the red line, yellow line, and blue line, respectively, with $p$ ranging in $[0.20,0.30]$. The orange and green areas represent the cases where $\cT^{\prime}_p$ is a CSPO, but $\QS^{+}(\cD_p, \cD_p) \circ T$ and $\QS^{+}(\cD_p \circ \sqrt{T}, \cD_p \circ \sqrt{T})$ admit non-stabilizerness.}
    \label{fig:qubit_QS_depo_PM} 
\end{figure}

We could notice that in each trajectory the effective channel is $\cT^{\prime}_p$, which will degenerate into a CSPO with a noise level $p > 0.261$. Under Q-SWITCH, we find that when $p < 0.275$, the latter two scenarios still result in an overall non-CSPO process, as shown in Fig.~\ref{fig:qubit_QS_depo_PM}.
The non-stabilizerness of $\cT^{\prime}_p$ can be characterized by its channel robustness of non-stabilizerness, i.e., $\cR_*(\cT^{\prime}_p)$ and its value is shown in the blue line with $p$ ranging in $[0.20,0.30]$.

We then compute the channel robustness of non-stabilizerness for $\QS^{+}(\cD_p, \cD_p) \circ T$ and $\QS^{+}(\cD_p \circ \sqrt{T}, \cD_p \circ \sqrt{T})$ for the qubit scenario, across a range of parameter values $p$ from $0.20$ to $0.30$ using the LP technique. Here the superscript `$+$' denotes the efficient channel when the measurement result of the control qubit is in `$+$'. The mathematical formulas for these processes can be found in Appendix~\ref{appendix:advantages_in_S+_S-}. From Fig.~\ref{fig:qubit_QS_depo_PM}, we could observe a notable shift in the threshold value of $p$ necessary to compromise the non-stabilizerness of the $T$ gate, moving from $0.261$ to $0.277$ when it traverses through $\QS^{+}(\cD_p, \cD_p) \circ T$ instead of two $p$-depolarizing channels applied sequentially. And for the case of $\QS^{+}(\cD_p \circ \sqrt{T}, \cD_p \circ \sqrt{T})$, the shift is moved from $0.261$ to $0.275$. Our results unveil the enhancement of the resilience of non-stabilizerness against noise-induced degradation. It coincides with the enhanced communication phenomenon through switched noisy channels from \cite{ebler2018enhanced}, while what is preserved here is the non-stabilizerness of the quantum gate.


\vspace{4mm}
\section{Conclusion}
In this work, we investigated the potential implications of indefinite causal networks on stabilizer operations, particularly CSPOs which are the largest set of operations providing no quantum advantage. We have introduced the magic resource capacity of a general quantum process and developed a quantitative method to estimate it. Based on our results, Q-SWTICH is a process that admits non-stabilizerness, and its non-stabilizerness generating power is relatively generic, by converting CSPO channels to non-CSPO ones.
Particularly, we have discovered that when we repeatedly use the same CSPO, even if each path has only the possibility to admit the same stabilizerness, their superposition appears a non-stabilizerness effect.

Our findings reveal the potential of indefinite causal order as a noise-robust resource to maintain the computational complexity of quantum computing, which classical resources lack the efficiency to simulate. It also provides valuable insights into the role of indefinite causal order in quantum resource theories. For future research, it would be intriguing to analyze more general indefinite causal networks from the perspective of quantum resource theories, such as communication~\cite{Kristj_nsson_2020}.
The question of whether all indefinite causal networks exhibit non-stabilizerness, the methodologies for quantifying the non-stabilizerness of indefinite causal networks, and the identification of input CSPOs that can maximally extract the non-stabilizerness of such supermaps are all intriguing avenues for analysis.
It will also be interesting to explore experimental demonstrations of our results via magic resource measures~\cite{Oliviero2022,Leone2022,Haug2023} on near-term quantum devices.
We hope that our work will inspire further investigations into the utilization of indefinite causal networks as a resource for unlocking the potential for quantum advantages.

\section{Acknowledgments}
We thank anonymous reviewers for their helpful comments and suggestions.
This work was partially supported by the National Key R\&D Program of China (No. 2024YFE0102500), the Guangdong Provincial Quantum Science Strategic Initiative (No. GDZX2303007), the Start-up Fund (No. G0101000151) from The Hong Kong University of Science and Technology (Guangzhou), and the Education Bureau of Guangzhou Municipality.

%

\appendix

\onecolumngrid
\vspace{2cm}

\section{Stabilizer formalism and discrete Wigner function}\label{appendix:preliminaries}
A resource-theoretic formalism can be adopted in which the free operations are quantum operations that can be efficiently simulated using a classical computer. We first recall the basic framework for the generalized stabilizer formalism. Let $\{\ket{j} \}_{j=0,\cdots,d-1}$ be a standard computational basis of a $d$-dimensional Hilbert space $\cH_d$. The unitary boost and shift operators $X, Z \in \mathscr{L}(\cH_d)$ are defined by~\cite{WWS19},
\begin{equation}
    X\ket{j} = \ket{j \oplus 1}, Z\ket{j} = w^j \ket{j}, w = e^{2\pi i/d},
\end{equation}
where $\oplus$ denotes addition modulo $d$. The \textit{discrete phase space} of a single $d$-level system is $\mathbb{Z}_d \times \mathbb{Z}_d$. At each point $\mathbf{u}=(a_1,a_2) \in \mathbb{Z}_d \times \mathbb{Z}_d$, the Heisenberg-Weyl operators in $\mathscr{L}(\cH_d)$ are given by
\begin{equation}
    T_{\mathbf{u}} = \left\{\begin{array}{cc}
    i Z^{a_1}X^{a_2},  &  d=2\\
    \tau^{-a_1 a_2} Z^{a_1} X^{a_2},  &  \text{ odd } d
    \end{array}\right.
\end{equation}
where $\tau=e^{(d+1) \pi i / d}$. The case of non-prime dimension can be understood as a tensor product of Hilbert spaces, each having a prime dimension. The Clifford group $\cC_d$ consists of unitary operators that map $T_{\mathbf{u}}$ to $T_{\mathbf{u}'}$ under unitary conjugation up to phases, i.e.,
\begin{equation}
    \cC_d = \{U: \forall \mathbf{u}, \exists \phi, \mathbf{u}', \text{ s.t. } UT_{\mathbf{u}}U^\dagger = e^{i\phi}T_{\mathbf{u}'}\}.
\end{equation}
The set of all pure stabilizer states in $\cH_d$ is then defined as $\{ U \ket{0}: U \in \mathcal{C}_d \}$. The stabilizer states ($\STAB$) are defined to be the convex hull of pure stabilizer states, and a state $\rho$ is called a magic state (or non-stabilizer state) if $\rho$ is not a stabilizer state. We denote $\STAB_n$ as the set of all $n$-qubit stabilizer states.


For each point $\mathbf{u} \in \mathbb{Z}_d \times \mathbb{Z}_d$ in the discrete phase space with odd prime dimension $d$, there is a phase-space point operator $A_{\mathbf{u}}$ defined as follows.
\begin{equation}
A_\mathbf{0}=\frac{1}{d} \sum_{\mathbf{u}} T_{\mathbf{u}}, \quad A_{\mathbf{u}}=T_{\mathbf{u}} A_\mathbf{0} T_{\mathbf{u}}^{\dagger}.
\end{equation} 
The discrete Wigner function of~\cite{WOOTTERS19871,Gross_2006a,Gross_2006b} a state $\rho$ at the point $\mathbf{u}$ is given by
\begin{equation}\label{Eq:wigner_f}
W_\rho(\mathbf{u}):=\frac{1}{d} \operatorname{Tr}\left[A_{\mathbf{u}} \rho\right].
\end{equation}
More generally, for a Hermitian operator $O$ acting on a space of dimension $d$, its discrete Wigner function is $W_O(\mathbf{u}) = \frac{1}{d}\text{Tr}[A_\mathbf{u}O]$. For the case of $O$ being a measurement operator $E$, we define the discrete Wigner function as:
\begin{align}
    W(E|\mathbf{u}) = \tr[EA_\mathbf{u}]
\end{align}
A set of all Hermitian operators with non-negative Wigner functions is defined as:
\begin{equation}
    \hat{\mathcal{W}}_+ \coloneqq \{V: \forall \mathbf{u}, W_V(\mathbf{u}) \geq 0 \}.
\end{equation}
In particular, the set of quantum states with non-negative Wigner function is denoted by:
\begin{equation}
    \mathcal{W}_+ \coloneqq \{\rho : \forall \mathbf{u}, W_{\rho}(\mathbf{u}) \geq 0, \rho \geq 0, \tr(\rho) = 1 \}.
\end{equation}
There are several useful properties of the set $\{A_\mathbf{u}\}$ as follows:
\begin{enumerate}
    \item $A_\mathbf{u}$ is Hermitian;
    \item $\sum_\mathbf{u} A_\mathbf{u}/d = \mathbb{I}$;
    \item $\text{Tr}[A_\mathbf{u}A_{u'}] = d \delta(u, u')$;
    \item $\text{Tr}[A_\mathbf{u}] = 1$;
    \item $X = \sum_\mathbf{u} W_{X}(\mathbf{u})A_\mathbf{u}$;
    \item $\{ A_\mathbf{u}\}_\mathbf{u} = \{ {A_\mathbf{u}}^T\}_\mathbf{u}$.  
\end{enumerate}

We denote the set of quantum states that have positive Wigner functions as $\cW_{+}$, i.e., 
\begin{equation}
    \cW_{+} := \{\rho: \rho \in \mathscr{D}(\cH_d), \forall \ \mathbf{u}, W_{\rho}(\mathbf{u})\geq 0\}.
\end{equation}
A magic monotone called mana~\cite{Veitch2014} of a quantum state $\rho$ is defined as
\begin{equation}
    \cM(\rho) := \log\left(\sum_{\mathbf{u}}|W_{\rho}(\mathbf{u})|\right).
\end{equation}
We have $\cM(\rho)\geq 0$ and $\cM(\rho) = 0$ if and only if $\cM(\rho)\in \cW_{+}$.
As a generalization of the Gottesman-Knill theorem, a quantum circuit where the initial state and all the following quantum operations have positive discrete Wigner functions can be classically simulated~\cite{Mari2012, Veitch2012}. Thus, for the QRT of magic states in odd prime dimensions, the free states can be chosen as $\cW_{+}$, and the free operations are those that completely preserve the positivity of Wigner functions~\cite{WW19}.
\begin{definition}[Completely PWP operation]
A Hermiticity-preserving linear map $\Pi$ is called completely positive Wigner preserving (CPWP) if, for any system $R$ with odd dimension, the following holds 
\begin{equation}
    \forall \rho_{RA}\in \cW_{+},  (\mathrm{id}_R\ox \Pi_{A\rightarrow B}) \rho_{RA} \in \cW_{+}.
\end{equation}
\end{definition}
To further explore the CPWP properties of quantum channels, the discrete Wigner function of a given quantum channel $\cN_{A \rightarrow B}$ is given by
\begin{align}\label{def:DWF_channel}
    \mathcal{W}_{\cN}(\mathbf{v}|\mathbf{u}) \coloneqq \frac{1}{d_B} \tr[((A_A^\mathbf{u})^T\otimes A_B^\mathbf{v}) J_{\cN}] = \frac{1}{d_B} \tr(A_B^\mathbf{v} \cN(A_A^\mathbf{u})).
\end{align}

\begin{definition}[Mana of a quantum channel]
The mana of a quantum channel $\cN_{A \rightarrow B}$ is defined as:
\begin{align}\label{def:Mana_channel}
    \mathcal{M}(\cN_{A \rightarrow B}) \coloneqq \log \max_\mathbf{u} \| \cN_{A \rightarrow B}(A_A^\mathbf{u})\|_{W,1} = \log \max_\mathbf{u} \sum_v |W_{\cN}(\mathbf{v}|\mathbf{u})|,
\end{align}
where $\|V \|_{W,1} \coloneqq \sum_\mathbf{u} |W_V(\mathbf{u})| = \sum_\mathbf{u} |\tr[A_{\mathbf{u}}V]/d|$ is the Wigner trace norm of an Hermitian operator $V$.     
\end{definition}
It is shown that $\cM(\cN_{A\rightarrow B}) \geq 0$ and $\cM(\cN_{A\rightarrow B}) = 0$ if and only if $\cN_{A\rightarrow B}\in \CPWP$~\cite{WWS19}. Also, it is proved that a quantum channel $\cN_{A\rightarrow B}$ is CPWP if and only if the discrete Wigner functions of $J_{\cN}$ are non-negative~\cite{WWS19}.

\section{Lower bound on the magic resource capacity of Q-SWTICH}
\label{appendix:magic_IDC}

In the following, we consider an $n$-qubit quantum switch $\QS_n$ as follows. Let $\ket{+} = 1/\sqrt{2}(\ket{0} + \ket{1})$ as the control qubit state for a $\QS$ on which the measurement is in the Fourier basis $\{\ket{+},\ket{-}\}$. The two input channels of $\QS_n$ are $n$-qubit quantum channels. Then we denote by $\QS_n^+$ the completely positive and trace-non-increasing (CPTN) map corresponding to the measurement operator $\ketbra{+}{+}$, and by $\QS_n^-$ the CPTN corresponding to the measurement operator $\ketbra{-}{-}$.  Based on these notations, we can obtain the following lower bound on the magic resource capacity of the quantum switch as introduced in the main text.
\begin{equation}
    \widetilde{\cC}'\big(\QS_n\big) := \log \max_{\cN_1,\cN_2} \bigg\{\cR\Big((\QS_n^+(\cN_1, \cN_2)\ox \mathds{1}_n)(\psi_{2n})\Big) + \cR\Big((\QS_n^-(\cN_1, \cN_2)\ox \mathds{1}_n)(\psi_{2n})\Big) \,:\, \psi_{2n}\in \STAB_{2n} \bigg\},
\end{equation}
where $\cN_1,\cN_2$ ranges over all CSPO channels. It is the lower bound of the magic resource capacity of Q-SWITCH since $\cR(\cdot)$ is convex-linear for quantum-classical states~\cite{gour2024resources} and $\cR(\ketbra{\psi}{\psi}\ox \sigma) = \cR(\sigma)$ if $\ket{\psi}$ is a stabilizer state. This quantity characterizes the average maximum magic resource that can be generated by inserting two quantum channels in the Q-SWITCH and inputting a stabilizer state. $(\QS_n^+(\cN_1, \cN_2)\ox \mathds{1}_n)(\psi_{2n})$ is the unnormalized post-measurement state when the measurement result is `$+$' on the control state and $(\QS_n^-(\cN_1, \cN_2)\ox \mathds{1}_n)(\psi_{2n})$ is the unnormalized post-measurement state when the measurement result is `$-$'.

Given two quantum channels $\cN_1,\cN_2$, denote the Choi operator of $\QS_n^+(\cN_1,\cN_2)$ as $S_n^+(N_1,N_2)$ where $N_1, N_2$ are the Choi operators of $\cN_1,\cN_2$, respectively. We note that maximally entangled states are stabilizer states, and hence have 
\begin{equation}
    \max_{\psi_{2n}}\cR\Big(\QS_n^+(\cN_1, \cN_2)\ox \mathds{1}_n)(\psi_{2n})\Big) \geq \cR(S_n^+(N_1,N_2))/2^{n} = \Big(\tr\big(S_n^+(N_1,N_2)\big) + 2\eta\Big)/2^n,
\end{equation}
where 
\begin{equation}
\begin{aligned}
\label{Eq: Magic_CPTN}
    \eta := \min& \;\; \tr M\\
    &\;\; {\rm s.t. } \;\; c_i, m_i \geq 0, \forall i,\\
    &\;\; S_n^+(N_1,N_2) + M = \sum_{i}c_i \ketbra{\psi_i}{\psi_i}\\
    &\;\; M = \sum_{i}m_i \ketbra{\psi_i}{\psi_i},
\end{aligned}
\end{equation}
where $\{\ket{\psi_i}\}_i$ is the set of all $2n$-qubit pure stabilizer states. The Lagrange function of this primal SDP problem can be written as
\begin{equation}
\begin{aligned}
    L &= \tr M + \tr\bigg[V\bigg(S_n^+(N_1,N_2) + M - \sum_{i}c_i \ketbra{\psi_i}{\psi_i}\bigg)\bigg] + \tr \bigg[ W\bigg(M -\sum_i m_i\ketbra{\psi_i}{\psi_i}\bigg)\bigg]\\
    &= \tr[M(I + V + W)] - \sum_{i}c_i \tr(V\ketbra{\psi_i}{\psi_i}) - \sum_{i}m_i \tr(W\ketbra{\psi_i}{\psi_i}) + \tr(VS_n^+(N_1,N_2)),
\end{aligned}
\end{equation}
where $V,W$ are dual variables. Thus, the dual problem can be written as 
\begin{equation}\label{Eq:dualSDP}
\begin{aligned}
    \max & \;\; \tr (VS_n^+(N_1,N_2))\\
    {\rm s.t. } &\;\; I+V+W \geq 0,\\
    &\;\; \bra{\psi_i}V\ket{\psi_i} \leq 0,\bra{\psi_i}W\ket{\psi_i} \leq 0,\forall i, \\
\end{aligned}
\end{equation}

Based on the above, the lower bound on the magic resource capacity of the quantum switch can be expressed as the following optimization problem
\begin{equation}\label{Eq:magic_cap_opt}
\begin{aligned}
   \widetilde{\cC}'\big(\QS_n\big) \geq \log \Big[ 1 + \frac{1}{2^{n-1}}\cdot \max & \; \tr (V_1S_n^+(N_1,N_2)) + \tr (V_2S_n^-(N_1,N_2)) \Big]\\
    {\rm s.t. } &\;\; a_i, b_i \geq 0, \forall i,\\
    &\;\; I+V_1+W_1 \geq 0,\; I+V_2+W_2 \geq 0,\\
    &\;\; \bra{\psi_i}V_1\ket{\psi_i} \leq 0,\; \bra{\psi_i}V_2\ket{\psi_i} \leq 0, \\
    &\;\; \bra{\psi_i}W_1\ket{\psi_i} \leq 0, \;\bra{\psi_i}W_2\ket{\psi_i} \leq 0, \forall i, \\
    &\;\; \sum_{i} a_i \ketbra{\psi_i}{\psi_i} = N_1,\; \sum_{i} b_i \ketbra{\psi_i}{\psi_i} = N_2,\\
    &\;\; \tr_{B}N_1 = I_A, \, \tr_{B}N_2 = I_A.\\
\end{aligned}
\end{equation}
Notice that the optimization problem in Eq.~\eqref{Eq:magic_cap_opt} is not a SDP. However, we introduce the following two SDPs to help estimate the optimal value of Eq.~\eqref{Eq:magic_cap_opt}. 
\begin{equation}\label{Eq:seesawSDP1}
\begin{aligned}
&\underline{\textbf{Program 1}}\\
    \max & \;\; \tr \Big(V_1S_n^+(N_1,N_2)\Big) + \tr \Big(V_2S_n^-(N_1,N_2)\Big)\\
    {\rm s.t. } &\;\; I+V_1+W_1 \geq 0, \, I+V_2+W_2 \geq 0,\\
    &\;\; \bra{\psi_i}V_1\ket{\psi_i} \leq 0,\bra{\psi_i}V_2\ket{\psi_i} \leq 0,\forall i, \\
    &\;\; \bra{\psi_i}W_1\ket{\psi_i} \leq 0,\bra{\psi_i}W_2\ket{\psi_i} \leq 0,\forall i, 
\end{aligned}
\end{equation}

\begin{equation}\label{Eq:see_saw_sdp}
\begin{aligned}
&\underline{\textbf{Program 2}}\\
\max & \; \tr \Big(V_1S_n^+(N_1,N_2)\Big) + \tr \Big(V_2S_n^-(N_1,N_2)\Big)\\
    {\rm s.t. } &\;\; a_i\geq 0, \sum_{i} a_i \ketbra{\psi_i}{\psi_i} = N_1,\\
    &\;\; \tr_{B}N_1 = I_A,\\
    &\;\; I+V_1+W_1 \geq 0,\, I+V_2+W_2 \geq 0,\\
    &\;\; \bra{\psi_i}W_1\ket{\psi_i} \leq 0,\bra{\psi_i}W_2\ket{\psi_i} \leq 0,\forall i, 
\end{aligned}
\hspace{6mm}
\begin{aligned}
&\underline{\textbf{Program 3}}\\
\max & \; \tr \Big(V_1S_n^+(N_1,N_2)\Big) + \tr \Big(V_2S^-_n(N_1,N_2)\Big)\\
    {\rm s.t. } &\;\; b_i\geq 0, \sum_{i} b_i \ketbra{\psi_i}{\psi_i} = N_2,\\
    &\;\; \tr_{B}N_2 = I_A,\\
    &\;\; I+V_1+W_1 \geq 0,\, I+V_2+W_2 \geq 0,\\
    &\;\; \bra{\psi_i}W_1\ket{\psi_i} \leq 0,\bra{\psi_i}W_2\ket{\psi_i} \leq 0,\forall i, 
\end{aligned}
\end{equation}
where in Program 1, the inputs are $\{N_1,N_2\}$, and the variables for optimization are $\{V_1,V_2,W_1,W_2\}$; in Program 2, the inputs are $\{V_1,V_2,N_2\}$, and the variables for optimization are $\{N_1, W_1,W_2\}$; in Program 3, the inputs are $\{V_1,V_2,N_1\}$, and the variables for optimization are $\{N_2, W_1,W_2\}$.

\begin{algorithm}\label{alg:protocol_depo}
    \renewcommand{\algorithmcfname}{Algorithm}
    \SetKwInOut{Input}{Input}\SetKwInOut{Output}{Output}
    \caption{Estimation of the magic resource capacity of $\QS_n^+$ \label{algo:inv_depo}}
    \Input{Two quantum channels $\cN_1, \cN_2$; number of iterations $K$;
    }
    \Output{A lower bound of $\cC(\QS)$;}
    $\widetilde{\cN_1}\leftarrow \cN_1$ and $\widetilde{\cN_2}\leftarrow \cN_2$\;
    \For{$k$ from $1$ to $K$}{
        Input $\widetilde{\cN_1},\widetilde{\cN_2}$ to the Program 1 in Eq.~\eqref{Eq:seesawSDP1} and output the solution $\widetilde{V}_1, \widetilde{V}_2$ for the optimal value $r_0^{(k)}$\;
        Input $\widetilde{V}_1, \widetilde{V}_2, \widetilde{\cN_2}$ to the Program 2 in Eq.~\eqref{Eq:see_saw_sdp} and output the solution $\widetilde{\cN_1}$ for the optimal value $r_1^{(k)}$\;
        Input $\widetilde{V}_1, \widetilde{V}_2, \widetilde{\cN_1}$ to the Program 3 in Eq.~\eqref{Eq:see_saw_sdp} and output the solution $\widetilde{\cN_2}$ for the optimal value $r_2^{(k)}$\;
    }
    Return $r_2^{(K)}$\;
\end{algorithm}

The algorithm for calculating a lower bound on $\widetilde{\cC}'\big(\QS_n\big)$ is described in Algorithm~\ref{alg:protocol_depo} which is based on the standard see-saw procedure. This algorithm is not guaranteed to converge to the global maximum. However, given a sufficient number of interactions, we hope to explore the potential magic-state-generating power of a quantum switch.

\section{Generate magic states with CSPO and Q-SWITCH}\label{appendix:qutrit_case}
\begin{example}
Consider a qutrit-to-qutrit quantum channel $\cN_{p}$ whose Kraus operators are given as 
\begin{equation}
\begin{aligned}
    &K_0 = \sqrt{\frac{p}{3}}\zeta(\ketbra{0}{0} + \ketbra{0}{1} + \ketbra{0}{2}), K_1 = \sqrt{\frac{p}{3}}(\ketbra{1}{0} + \omega\ketbra{1}{1} + \omega^2\ketbra{1}{2}),\\
    &K_2 = \sqrt{\frac{p}{3}}\zeta(\ketbra{2}{0} + \omega^2\ketbra{1}{1} + \omega\ketbra{1}{2}), K_3 = \sqrt{1-p}TH,
\end{aligned}
\end{equation}
where $\omega = e^{2\pi i/3},\zeta = e^{2\pi i/9}$, $H$, $T:={\rm diag}(\zeta, 1, \zeta^{-1})$ are qutrit Hadamard gate and $T$ gate respectively. When $p\gtrsim 0.47$, although $\cN_{p}$ is CPWP, a magic state could be generated from $\QS(\cN_{p}, \cN_{p})$ acting on a stabilizer state.
\end{example}

We investigate a qutrit input target state $\ket{+}_{\rm tr} = (\ket{0}+\ket{1}+\ket{2})/\sqrt{3}$. Set the control qubit to $\ket{+}_c=(\ket{0}+\ket{1})/\sqrt{2}$ and measure it in the qubit Fourier basis $\{\ket{+}_c, \ket{-}_c\}$ at last. We note given a quantum state $\rho$, the mana of $\rho$, i.e., $\cM(\rho)$ is a faithful magic monotone that can be used to quantify the non-stabilizerness of the state. Consequently, we can calculate the mana of $\rho_{\QS,+}(p)$ and $\nonumber\rho_{\QS,-}(p)$ to determine whether the resulting output state is a magic state or not. Similarly, the mana of a quantum channel $\cN_p$, i.e., $\cM(\cN_p)$, is also a faithful magic monotone for quantifying the non-stabilizerness of the channel. We then utilize it to determine whether the input channel for the Q-SWITCH admits non-stabilizerness or not.

\begin{figure*}[t]
    \centering
    \includegraphics[width=0.55\linewidth]{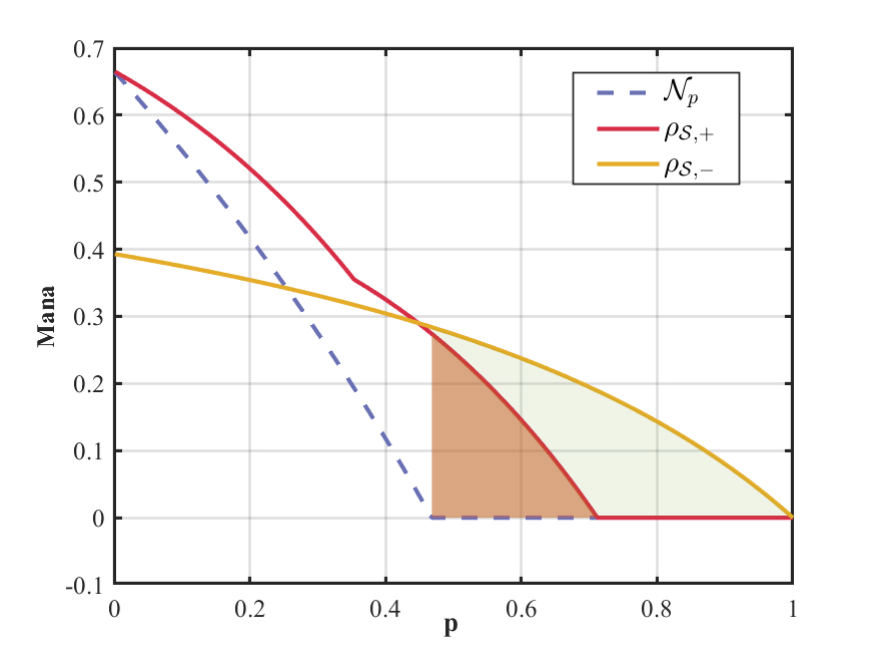}
    \caption{An example for Theorem \textcolor{blue}{1} in the qutrit scenario. The mana of $\rho_{\QS,+}(p)$, $\rho_{\QS,-}(p)$ are shown in the red line and the yellow line, respectively. The mana of $\cN_p$ is shown in the blue line, with $p$ ranging in $(0,1]$. The orange and green areas represent the cases where $\cN_p$ is a CSPO and $\rho_{\QS,+}(p)$ or $\rho_{\QS,-}(p)$ is a magic state.}
    \label{fig:qutrit_qs_seq}
\end{figure*}

In Fig.~\ref{fig:qutrit_qs_seq}, we present $\cM(\rho_{\QS,+}(p)), \cM(\rho_{\QS,-}(p))$, and $\cM(\cN_{p})$ respectively. For $p$ in the range $[0.4679, 0.7129]$, we obtain a magic state in the target system regardless of the measurement result is `$+$' or `$-$', even though the preparation of the state $\ket{+}$, the channel $\cN_{p}$ and the measurement on the control system are all stabilizer operations. Interestingly, when $p$ is in the range $[0.7129,1)$, we still will obtain a magic state when the measurement result is `$-$'.

\section{The mathematical formulas for \texorpdfstring{$\QS(\cD_p, \cD_p) \circ T$}{TEXT} and \texorpdfstring{$\QS(\cD_p \circ \sqrt{T}, \cD_p \circ \sqrt{T})$}{TEXT}}\label{appendix:advantages_in_S+_S-}
In this subsection, we give the detailed calculation for the process of $\QS(\cD_p, \cD_p) \circ T$ and $\QS(\cD_p \circ \sqrt{T}, \cD_p \circ \sqrt{T})$ to show how the $T$ gate is affected by noise in Q-SWITCH. The channel robustness of non-stabilizerness for the final process could then be calculated by the linear programming (LP) scheme developed by Seddon and Campbell~\cite{Seddon_2019}. 

The expression for $\QS(\cD_p, \cD_p)(\cdot)$ has been derived in the supplemental material of Ref.~\cite{ebler2018enhanced}. Here, we give a slightly adjusted version of the calculation and analyze the magic resource of the final process after measuring the control qubit in the Fourier basis.
When the control qubit is initialized to $\ket{+}_c$, a state $\rho$ after passing through two $p$-depolarizing channels with Q-SWITCH becomes:
\begin{align}
    \QS(\cD_p, \cD_p)(\rho\otimes \ket{+}\bra{+}_c ) 
    = \frac{p^2}{2d} I_c \otimes I + \ket{+}\bra{+}_c \otimes\Big(\frac{2p(1-p)}{d} I + (1-p)^2 \rho\Big) 
    + \frac{p^2}{2d^2}(\ket{0}\bra{1}_c + \ket{1}\bra{0}_c) \otimes \rho \, ,
\end{align}
where the calculation is based on the fact that $\{U_i\}$ forms an orthonormal basis for the set of $d \times d$ matrices, and $\sum_i \tr[U_i \rho]U_i^{\dagger} = d\rho$, $\sum_i U_i \rho U_i^\dagger = dI$.
If the control qubit is then measured in the Fourier basis and the measurement result is `$+$', the target state (unnormalized) becomes
\begin{align}
    \nonumber\bra{+} &\QS(\cD_p, \cD_p)(\rho\otimes \ket{+}\bra{+}_c ) \ket{+}\\
    \nonumber&= \frac{p^2}{2d}  I + \frac{2p(1-p)}{d} I + (1-p)^2 \rho + \frac{p^2}{2d^2}\rho \\
    &= \frac{4p-3p^2}{2d} I + (1-2p+p^2 + \frac{p^2}{2d^2}) \rho.
\end{align}
The expression for the output state when the measurement outcome is `$-$' can be obtained similarly. Based on this, we denote the whole map acting on the target input when the control qubit's measurement outcome is `$\pm$' as 
\begin{equation}\label{Eq:QS_pm}
\begin{aligned}
    \QS^{+}(\cD_p, \cD_p) (\cdot)
    =& \frac{2d^2-(d^2-1)p^2}{2d^2}\cD_{p'_+}(\cdot),\\
    \QS^{-}(\cD_p, \cD_p)(\cdot)
    =& \frac{(d^2-1)p^2}{2d^2} \cD_{p'_-}(\cdot) \, ,
\end{aligned}
\end{equation}
where $p'_+=\frac{d^2(4p-3p^2)}{2d^2-(d^2-1)p^2}$ and $p'_-=\frac{d^2}{d^2-1}$.
For $\QS^{+}(\cD_p, \cD_p)(\cdot)$, it is equivalent to passing through a $p'_+$-depolarizing channel. 
Here we show that $p'_+$ is smaller than $2p-p^2$, which is the noise sequentially passing through two $p$-depolarizing channels, for arbitrary $p$ and $d$:
\begin{align*}
    (2d^2-(d^2-1)p^2) \cdot (2p-p^2 - p'_+) &= (2d^2-(d^2-1)p^2)\cdot (2p-p^2) - d^2(4p-3p^2) \\
    &= 4d^2 p - 2d^2 p^2 - 2(d^2-1)p^3 + (d^2-1)p^4 - 4d^2 p + 3d^2 p^2 \\
    &= d^2 p^2 - 2(d^2-1)p^3 + (d^2-1)p^4\\
    &= p^2 \cdot \Big(d^2 -(d^2-1)p(2-p)\Big).
\end{align*}
Since in range $[0,1]$, $p(2-p)$ is monotone increasing, we substitute $p=1$ into the above equation and get
\begin{align*}
    (2d^2-(d^2-1)p^2) \cdot (2p-p^2 - p'_+) > d^2 -(d^2-1) > 0 \, .
\end{align*}

\begin{figure}[t]
    \centering
    \includegraphics[width=0.55\linewidth]{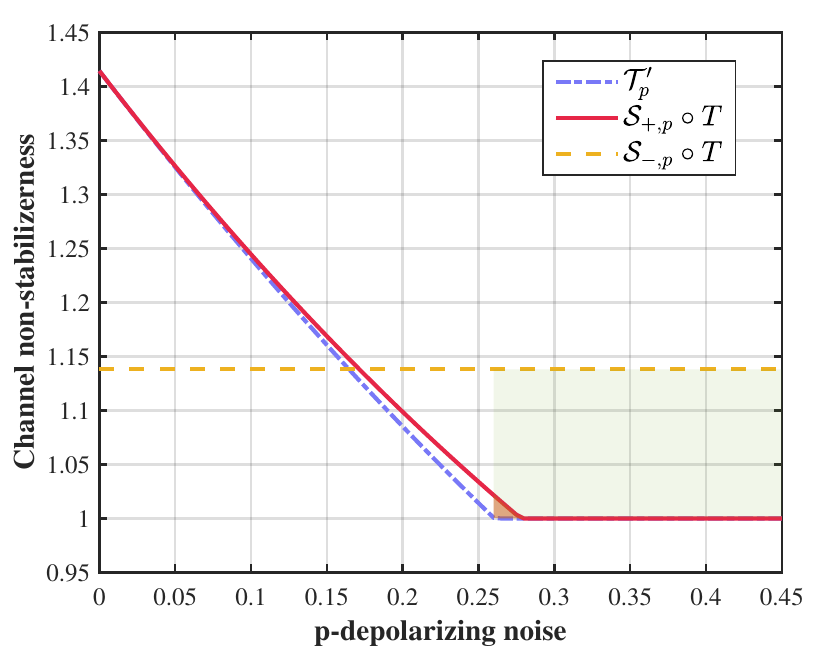}
    \caption{
    Comparison between the non-stabilizerness of $T$ gate after the depolarizing noise under Q-SWITCH and the non-stabilizerness of $T$ gate after two sequential depolarizing noises. This figure depicts the channel robustness of $\QS^{+}(\cD_p, \cD_p) \circ T$, $\QS^{-}(\cD_p, \cD_p) \circ T$ and $\cT^{\prime}_p$ in the red line, yellow line, and blue line, respectively, with $p$ ranging in $[0, 0.45]$. The orange and green areas represent the cases where $\cT^{\prime}_p$ is a CSPO, but $\QS^{+}(\cD_p, \cD_p) \circ T$ and $\QS^{-}(\cD_p, \cD_p) \circ T$ admit non-stabilizerness.} 
    \label{fig:qubit_QS_depo_PM2}
\end{figure}

For $\QS^{-}(\cD_p, \cD_p)(\cdot)$, it's worth noting that it resembles the form of a depolarizing channel, with $p'_-\geq 1$ and being independent of the parameter $p$. We compute the channel robustness of non-stabilizerness for $\QS^{+}(\cD_p, \cD_p) \circ T$, $\QS^{-}(\cD_p, \cD_p) \circ T$ for the qubit scenario, across a range of parameter values $p$ from $0$ to $0.45$ using an LP technique. The results are depicted in Fig.~\ref{fig:qubit_QS_depo_PM2}. We find that the channel robustness for $\QS^{-}(\cD_p, \cD_p) \circ T$ is always larger than $1$, and remains unaffected by the noise level.

For $\QS(\cD_p \circ \sqrt{T}, \cD_p \circ \sqrt{T})$, based on similar calculation, we could get 
\begin{align}
    \nonumber \QS(\cD_p \circ \sqrt{T}, \cD_p \circ \sqrt{T})(\rho\otimes \ket{+}\bra{+}_c ) 
    = &\quad \frac{p^2+2p(1-p)}{2d} I_c \otimes I \\
    \nonumber & + (1-p)^2 \ket{+}\bra{+}_c \otimes T \rho T^{\dagger}\\
    & + (\ket{0}\bra{1}_c + \ket{1}\bra{0}_c) \otimes \Big( \frac{p^2}{2d^2} \rho + \frac{p(1-p)}{2d}(\tr\Big[\sqrt{T}\rho\Big]\sqrt{T}^{\dagger} + \tr\Big[\sqrt{T}^{\dagger}\rho\Big]\sqrt{T}) \Big) \, .
\end{align}
After measuring the control qubit in the Fourier basis, the final process is given as
\begin{equation}
\begin{aligned}
    \QS^{+}(\cD_p \circ \sqrt{T}, \cD_p \circ \sqrt{T}) (\rho)
    =& \frac{p^2+2p(1-p)}{2d} I + (1-p)^2 T \rho T^{\dagger} + \frac{p^2}{2d^2} \rho + \frac{p(1-p)}{2d}(\tr\Big[\sqrt{T}\rho\Big]\sqrt{T}^{\dagger} + \tr\Big[\sqrt{T}^{\dagger}\rho\Big]\sqrt{T}) \, ,\\
    \QS^{-}(\cD_p \circ \sqrt{T}, \cD_p \circ \sqrt{T}) (\rho)
    =&  \frac{p^2+2p(1-p)}{2d} I - \frac{p^2}{2d^2} \rho - \frac{p(1-p)}{2d}(\tr\Big[\sqrt{T}\rho\Big]\sqrt{T}^{\dagger} + \tr\Big[\sqrt{T}^{\dagger}\rho\Big]\sqrt{T})\, .
\end{aligned}
\end{equation}
The channel robustness could then be calculated directly.


\begin{thebibliography}{61}%
\makeatletter
\providecommand \@ifxundefined [1]{%
 \@ifx{#1\undefined}
}%
\providecommand \@ifnum [1]{%
 \ifnum #1\expandafter \@firstoftwo
 \else \expandafter \@secondoftwo
 \fi
}%
\providecommand \@ifx [1]{%
 \ifx #1\expandafter \@firstoftwo
 \else \expandafter \@secondoftwo
 \fi
}%
\providecommand \natexlab [1]{#1}%
\providecommand \enquote  [1]{``#1''}%
\providecommand \bibnamefont  [1]{#1}%
\providecommand \bibfnamefont [1]{#1}%
\providecommand \citenamefont [1]{#1}%
\providecommand \href@noop [0]{\@secondoftwo}%
\providecommand \href [0]{\begingroup \@sanitize@url \@href}%
\providecommand \@href[1]{\@@startlink{#1}\@@href}%
\providecommand \@@href[1]{\endgroup#1\@@endlink}%
\providecommand \@sanitize@url [0]{\catcode `\\12\catcode `\$12\catcode
  `\&12\catcode `\#12\catcode `\^12\catcode `\_12\catcode `\%12\relax}%
\providecommand \@@startlink[1]{}%
\providecommand \@@endlink[0]{}%
\providecommand \url  [0]{\begingroup\@sanitize@url \@url }%
\providecommand \@url [1]{\endgroup\@href {#1}{\urlprefix }}%
\providecommand \urlprefix  [0]{URL }%
\providecommand \Eprint [0]{\href }%
\providecommand \doibase [0]{https://doi.org/}%
\providecommand \selectlanguage [0]{\@gobble}%
\providecommand \bibinfo  [0]{\@secondoftwo}%
\providecommand \bibfield  [0]{\@secondoftwo}%
\providecommand \translation [1]{[#1]}%
\providecommand \BibitemOpen [0]{}%
\providecommand \bibitemStop [0]{}%
\providecommand \bibitemNoStop [0]{.\EOS\space}%
\providecommand \EOS [0]{\spacefactor3000\relax}%
\providecommand \BibitemShut  [1]{\csname bibitem#1\endcsname}%
\let\auto@bib@innerbib\@empty
\bibitem [{\citenamefont {Shor}(1997)}]{shor1997faulttolerant}%
  \BibitemOpen
  \bibfield  {author} {\bibinfo {author} {\bibfnamefont {P.~W.}\ \bibnamefont
  {Shor}},\ }\href@noop {} {\bibinfo {title} {Fault-tolerant quantum
  computation}} (\bibinfo {year} {1997}),\ \Eprint
  {https://arxiv.org/abs/quant-ph/9605011} {arXiv:quant-ph/9605011 [quant-ph]}
  \BibitemShut {NoStop}%
\bibitem [{\citenamefont {Renner}\ and\ \citenamefont
  {Brukner}(2022)}]{renner2022computational}%
  \BibitemOpen
  \bibfield  {author} {\bibinfo {author} {\bibfnamefont {M.~J.}\ \bibnamefont
  {Renner}}\ and\ \bibinfo {author} {\bibfnamefont {{\v{C}}.}~\bibnamefont
  {Brukner}},\ }\bibfield  {title} {\bibinfo {title} {Computational advantage
  from a quantum superposition of qubit gate orders},\ }\href
  {https://journals.aps.org/prl/abstract/10.1103/PhysRevLett.128.230503}
  {\bibfield  {journal} {\bibinfo  {journal} {Physical Review Letters}\
  }\textbf {\bibinfo {volume} {128}},\ \bibinfo {pages} {230503} (\bibinfo
  {year} {2022})}\BibitemShut {NoStop}%
\bibitem [{\citenamefont {Wu}\ \emph {et~al.}(2021)\citenamefont {Wu},
  \citenamefont {Bao}, \citenamefont {Cao}, \citenamefont {Chen}, \citenamefont
  {Chen}, \citenamefont {Chen}, \citenamefont {Chung}, \citenamefont {Deng},
  \citenamefont {Du}, \citenamefont {Fan} \emph {et~al.}}]{Wu2021strong}%
  \BibitemOpen
  \bibfield  {author} {\bibinfo {author} {\bibfnamefont {Y.}~\bibnamefont
  {Wu}}, \bibinfo {author} {\bibfnamefont {W.-S.}\ \bibnamefont {Bao}},
  \bibinfo {author} {\bibfnamefont {S.}~\bibnamefont {Cao}}, \bibinfo {author}
  {\bibfnamefont {F.}~\bibnamefont {Chen}}, \bibinfo {author} {\bibfnamefont
  {M.-C.}\ \bibnamefont {Chen}}, \bibinfo {author} {\bibfnamefont
  {X.}~\bibnamefont {Chen}}, \bibinfo {author} {\bibfnamefont {T.-H.}\
  \bibnamefont {Chung}}, \bibinfo {author} {\bibfnamefont {H.}~\bibnamefont
  {Deng}}, \bibinfo {author} {\bibfnamefont {Y.}~\bibnamefont {Du}}, \bibinfo
  {author} {\bibfnamefont {D.}~\bibnamefont {Fan}}, \emph {et~al.},\ }\bibfield
   {title} {\bibinfo {title} {Strong quantum computational advantage using a
  superconducting quantum processor},\ }\href
  {https://journals.aps.org/prl/abstract/10.1103/PhysRevLett.127.180501}
  {\bibfield  {journal} {\bibinfo  {journal} {Physical Review Letters}\
  }\textbf {\bibinfo {volume} {127}},\ \bibinfo {pages} {180501} (\bibinfo
  {year} {2021})}\BibitemShut {NoStop}%
\bibitem [{\citenamefont {Knill}(2005)}]{knill2005quantum}%
  \BibitemOpen
  \bibfield  {author} {\bibinfo {author} {\bibfnamefont {E.}~\bibnamefont
  {Knill}},\ }\bibfield  {title} {\bibinfo {title} {Quantum computing with
  realistically noisy devices},\ }\href {http://dx.doi.org/10.1038/nature03350}
  {\bibfield  {journal} {\bibinfo  {journal} {Nature}\ }\textbf {\bibinfo
  {volume} {434}},\ \bibinfo {pages} {39} (\bibinfo {year} {2005})}\BibitemShut
  {NoStop}%
\bibitem [{\citenamefont {Shor}(1999)}]{Shor1999polynomial}%
  \BibitemOpen
  \bibfield  {author} {\bibinfo {author} {\bibfnamefont {P.~W.}\ \bibnamefont
  {Shor}},\ }\bibfield  {title} {\bibinfo {title} {Polynomial-time algorithms
  for prime factorization and discrete logarithms on a quantum computer},\
  }\href {http://dx.doi.org/10.1137/S0097539795293172} {\bibfield  {journal}
  {\bibinfo  {journal} {SIAM review}\ }\textbf {\bibinfo {volume} {41}},\
  \bibinfo {pages} {303} (\bibinfo {year} {1999})}\BibitemShut {NoStop}%
\bibitem [{\citenamefont {Arute}\ \emph {et~al.}(2019)\citenamefont {Arute},
  \citenamefont {Arya}, \citenamefont {Babbush}, \citenamefont {Bacon},
  \citenamefont {Bardin}, \citenamefont {Barends}, \citenamefont {Biswas},
  \citenamefont {Boixo}, \citenamefont {Brandao}, \citenamefont {Buell},
  \citenamefont {Burkett}, \citenamefont {Chen}, \citenamefont {Chen},
  \citenamefont {Chiaro}, \citenamefont {Collins}, \citenamefont {Courtney},
  \citenamefont {Dunsworth}, \citenamefont {Farhi}, \citenamefont {Foxen},
  \citenamefont {Fowler}, \citenamefont {Gidney}, \citenamefont {Giustina},
  \citenamefont {Graff}, \citenamefont {Guerin}, \citenamefont {Habegger},
  \citenamefont {Harrigan}, \citenamefont {Hartmann}, \citenamefont {Ho},
  \citenamefont {Hoffmann}, \citenamefont {Huang}, \citenamefont {Humble},
  \citenamefont {Isakov}, \citenamefont {Jeffrey}, \citenamefont {Jiang},
  \citenamefont {Kafri}, \citenamefont {Kechedzhi}, \citenamefont {Kelly},
  \citenamefont {Klimov}, \citenamefont {Knysh}, \citenamefont {Korotkov},
  \citenamefont {Kostritsa}, \citenamefont {Landhuis}, \citenamefont
  {Lindmark}, \citenamefont {Lucero}, \citenamefont {Lyakh}, \citenamefont
  {Mandr{\`{a}}}, \citenamefont {McClean}, \citenamefont {McEwen},
  \citenamefont {Megrant}, \citenamefont {Mi}, \citenamefont {Michielsen},
  \citenamefont {Mohseni}, \citenamefont {Mutus}, \citenamefont {Naaman},
  \citenamefont {Neeley}, \citenamefont {Neill}, \citenamefont {Niu},
  \citenamefont {Ostby}, \citenamefont {Petukhov}, \citenamefont {Platt},
  \citenamefont {Quintana}, \citenamefont {Rieffel}, \citenamefont {Roushan},
  \citenamefont {Rubin}, \citenamefont {Sank}, \citenamefont {Satzinger},
  \citenamefont {Smelyanskiy}, \citenamefont {Sung}, \citenamefont
  {Trevithick}, \citenamefont {Vainsencher}, \citenamefont {Villalonga},
  \citenamefont {White}, \citenamefont {Yao}, \citenamefont {Yeh},
  \citenamefont {Zalcman}, \citenamefont {Neven},\ and\ \citenamefont
  {Martinis}}]{Arute2019}%
  \BibitemOpen
  \bibfield  {author} {\bibinfo {author} {\bibfnamefont {F.}~\bibnamefont
  {Arute}}, \bibinfo {author} {\bibfnamefont {K.}~\bibnamefont {Arya}},
  \bibinfo {author} {\bibfnamefont {R.}~\bibnamefont {Babbush}}, \bibinfo
  {author} {\bibfnamefont {D.}~\bibnamefont {Bacon}}, \bibinfo {author}
  {\bibfnamefont {J.~C.}\ \bibnamefont {Bardin}}, \bibinfo {author}
  {\bibfnamefont {R.}~\bibnamefont {Barends}}, \bibinfo {author} {\bibfnamefont
  {R.}~\bibnamefont {Biswas}}, \bibinfo {author} {\bibfnamefont
  {S.}~\bibnamefont {Boixo}}, \bibinfo {author} {\bibfnamefont {F.~G. S.~L.}\
  \bibnamefont {Brandao}}, \bibinfo {author} {\bibfnamefont {D.~A.}\
  \bibnamefont {Buell}}, \bibinfo {author} {\bibfnamefont {B.}~\bibnamefont
  {Burkett}}, \bibinfo {author} {\bibfnamefont {Y.}~\bibnamefont {Chen}},
  \bibinfo {author} {\bibfnamefont {Z.}~\bibnamefont {Chen}}, \bibinfo {author}
  {\bibfnamefont {B.}~\bibnamefont {Chiaro}}, \bibinfo {author} {\bibfnamefont
  {R.}~\bibnamefont {Collins}}, \bibinfo {author} {\bibfnamefont
  {W.}~\bibnamefont {Courtney}}, \bibinfo {author} {\bibfnamefont
  {A.}~\bibnamefont {Dunsworth}}, \bibinfo {author} {\bibfnamefont
  {E.}~\bibnamefont {Farhi}}, \bibinfo {author} {\bibfnamefont
  {B.}~\bibnamefont {Foxen}}, \bibinfo {author} {\bibfnamefont
  {A.}~\bibnamefont {Fowler}}, \bibinfo {author} {\bibfnamefont
  {C.}~\bibnamefont {Gidney}}, \bibinfo {author} {\bibfnamefont
  {M.}~\bibnamefont {Giustina}}, \bibinfo {author} {\bibfnamefont
  {R.}~\bibnamefont {Graff}}, \bibinfo {author} {\bibfnamefont
  {K.}~\bibnamefont {Guerin}}, \bibinfo {author} {\bibfnamefont
  {S.}~\bibnamefont {Habegger}}, \bibinfo {author} {\bibfnamefont {M.~P.}\
  \bibnamefont {Harrigan}}, \bibinfo {author} {\bibfnamefont {M.~J.}\
  \bibnamefont {Hartmann}}, \bibinfo {author} {\bibfnamefont {A.}~\bibnamefont
  {Ho}}, \bibinfo {author} {\bibfnamefont {M.}~\bibnamefont {Hoffmann}},
  \bibinfo {author} {\bibfnamefont {T.}~\bibnamefont {Huang}}, \bibinfo
  {author} {\bibfnamefont {T.~S.}\ \bibnamefont {Humble}}, \bibinfo {author}
  {\bibfnamefont {S.~V.}\ \bibnamefont {Isakov}}, \bibinfo {author}
  {\bibfnamefont {E.}~\bibnamefont {Jeffrey}}, \bibinfo {author} {\bibfnamefont
  {Z.}~\bibnamefont {Jiang}}, \bibinfo {author} {\bibfnamefont
  {D.}~\bibnamefont {Kafri}}, \bibinfo {author} {\bibfnamefont
  {K.}~\bibnamefont {Kechedzhi}}, \bibinfo {author} {\bibfnamefont
  {J.}~\bibnamefont {Kelly}}, \bibinfo {author} {\bibfnamefont {P.~V.}\
  \bibnamefont {Klimov}}, \bibinfo {author} {\bibfnamefont {S.}~\bibnamefont
  {Knysh}}, \bibinfo {author} {\bibfnamefont {A.}~\bibnamefont {Korotkov}},
  \bibinfo {author} {\bibfnamefont {F.}~\bibnamefont {Kostritsa}}, \bibinfo
  {author} {\bibfnamefont {D.}~\bibnamefont {Landhuis}}, \bibinfo {author}
  {\bibfnamefont {M.}~\bibnamefont {Lindmark}}, \bibinfo {author}
  {\bibfnamefont {E.}~\bibnamefont {Lucero}}, \bibinfo {author} {\bibfnamefont
  {D.}~\bibnamefont {Lyakh}}, \bibinfo {author} {\bibfnamefont
  {S.}~\bibnamefont {Mandr{\`{a}}}}, \bibinfo {author} {\bibfnamefont {J.~R.}\
  \bibnamefont {McClean}}, \bibinfo {author} {\bibfnamefont {M.}~\bibnamefont
  {McEwen}}, \bibinfo {author} {\bibfnamefont {A.}~\bibnamefont {Megrant}},
  \bibinfo {author} {\bibfnamefont {X.}~\bibnamefont {Mi}}, \bibinfo {author}
  {\bibfnamefont {K.}~\bibnamefont {Michielsen}}, \bibinfo {author}
  {\bibfnamefont {M.}~\bibnamefont {Mohseni}}, \bibinfo {author} {\bibfnamefont
  {J.}~\bibnamefont {Mutus}}, \bibinfo {author} {\bibfnamefont
  {O.}~\bibnamefont {Naaman}}, \bibinfo {author} {\bibfnamefont
  {M.}~\bibnamefont {Neeley}}, \bibinfo {author} {\bibfnamefont
  {C.}~\bibnamefont {Neill}}, \bibinfo {author} {\bibfnamefont {M.~Y.}\
  \bibnamefont {Niu}}, \bibinfo {author} {\bibfnamefont {E.}~\bibnamefont
  {Ostby}}, \bibinfo {author} {\bibfnamefont {A.}~\bibnamefont {Petukhov}},
  \bibinfo {author} {\bibfnamefont {J.~C.}\ \bibnamefont {Platt}}, \bibinfo
  {author} {\bibfnamefont {C.}~\bibnamefont {Quintana}}, \bibinfo {author}
  {\bibfnamefont {E.~G.}\ \bibnamefont {Rieffel}}, \bibinfo {author}
  {\bibfnamefont {P.}~\bibnamefont {Roushan}}, \bibinfo {author} {\bibfnamefont
  {N.~C.}\ \bibnamefont {Rubin}}, \bibinfo {author} {\bibfnamefont
  {D.}~\bibnamefont {Sank}}, \bibinfo {author} {\bibfnamefont {K.~J.}\
  \bibnamefont {Satzinger}}, \bibinfo {author} {\bibfnamefont {V.}~\bibnamefont
  {Smelyanskiy}}, \bibinfo {author} {\bibfnamefont {K.~J.}\ \bibnamefont
  {Sung}}, \bibinfo {author} {\bibfnamefont {M.~D.}\ \bibnamefont
  {Trevithick}}, \bibinfo {author} {\bibfnamefont {A.}~\bibnamefont
  {Vainsencher}}, \bibinfo {author} {\bibfnamefont {B.}~\bibnamefont
  {Villalonga}}, \bibinfo {author} {\bibfnamefont {T.}~\bibnamefont {White}},
  \bibinfo {author} {\bibfnamefont {Z.~J.}\ \bibnamefont {Yao}}, \bibinfo
  {author} {\bibfnamefont {P.}~\bibnamefont {Yeh}}, \bibinfo {author}
  {\bibfnamefont {A.}~\bibnamefont {Zalcman}}, \bibinfo {author} {\bibfnamefont
  {H.}~\bibnamefont {Neven}},\ and\ \bibinfo {author} {\bibfnamefont {J.~M.}\
  \bibnamefont {Martinis}},\ }\bibfield  {title} {\bibinfo {title} {{Quantum
  supremacy using a programmable superconducting processor}},\ }\href
  {https://doi.org/10.1038/s41586-019-1666-5} {\bibfield  {journal} {\bibinfo
  {journal} {Nature}\ }\textbf {\bibinfo {volume} {574}},\ \bibinfo {pages}
  {505} (\bibinfo {year} {2019})},\ \Eprint {https://arxiv.org/abs/1910.11333}
  {arXiv:1910.11333} \BibitemShut {NoStop}%
\bibitem [{\citenamefont {Zhong}\ \emph {et~al.}(2020)\citenamefont {Zhong},
  \citenamefont {Wang}, \citenamefont {Deng}, \citenamefont {Chen},
  \citenamefont {Peng}, \citenamefont {Luo}, \citenamefont {Qin}, \citenamefont
  {Wu}, \citenamefont {Ding}, \citenamefont {Hu} \emph
  {et~al.}}]{Zhong2020quantum}%
  \BibitemOpen
  \bibfield  {author} {\bibinfo {author} {\bibfnamefont {H.-S.}\ \bibnamefont
  {Zhong}}, \bibinfo {author} {\bibfnamefont {H.}~\bibnamefont {Wang}},
  \bibinfo {author} {\bibfnamefont {Y.-H.}\ \bibnamefont {Deng}}, \bibinfo
  {author} {\bibfnamefont {M.-C.}\ \bibnamefont {Chen}}, \bibinfo {author}
  {\bibfnamefont {L.-C.}\ \bibnamefont {Peng}}, \bibinfo {author}
  {\bibfnamefont {Y.-H.}\ \bibnamefont {Luo}}, \bibinfo {author} {\bibfnamefont
  {J.}~\bibnamefont {Qin}}, \bibinfo {author} {\bibfnamefont {D.}~\bibnamefont
  {Wu}}, \bibinfo {author} {\bibfnamefont {X.}~\bibnamefont {Ding}}, \bibinfo
  {author} {\bibfnamefont {Y.}~\bibnamefont {Hu}}, \emph {et~al.},\ }\bibfield
  {title} {\bibinfo {title} {Quantum computational advantage using photons},\
  }\href {http://dx.doi.org/10.1126/science.abe8770} {\bibfield  {journal}
  {\bibinfo  {journal} {Science}\ }\textbf {\bibinfo {volume} {370}},\ \bibinfo
  {pages} {1460} (\bibinfo {year} {2020})}\BibitemShut {NoStop}%
\bibitem [{\citenamefont {Chiribella}\ \emph {et~al.}(2013)\citenamefont
  {Chiribella}, \citenamefont {D'Ariano}, \citenamefont {Perinotti},\ and\
  \citenamefont {Valiron}}]{Chiribella_2013}%
  \BibitemOpen
  \bibfield  {author} {\bibinfo {author} {\bibfnamefont {G.}~\bibnamefont
  {Chiribella}}, \bibinfo {author} {\bibfnamefont {G.~M.}\ \bibnamefont
  {D'Ariano}}, \bibinfo {author} {\bibfnamefont {P.}~\bibnamefont
  {Perinotti}},\ and\ \bibinfo {author} {\bibfnamefont {B.}~\bibnamefont
  {Valiron}},\ }\bibfield  {title} {\bibinfo {title} {Quantum computations
  without definite causal structure},\ }\bibfield  {journal} {\bibinfo
  {journal} {Physical Review A}\ }\textbf {\bibinfo {volume} {88}},\ \href
  {https://doi.org/10.1103/physreva.88.022318} {10.1103/physreva.88.022318}
  (\bibinfo {year} {2013})\BibitemShut {NoStop}%
\bibitem [{\citenamefont {Oreshkov}\ \emph {et~al.}(2012)\citenamefont
  {Oreshkov}, \citenamefont {Costa},\ and\ \citenamefont
  {Brukner}}]{Oreshkov_2012}%
  \BibitemOpen
  \bibfield  {author} {\bibinfo {author} {\bibfnamefont {O.}~\bibnamefont
  {Oreshkov}}, \bibinfo {author} {\bibfnamefont {F.}~\bibnamefont {Costa}},\
  and\ \bibinfo {author} {\bibfnamefont {{\v{C}}.}~\bibnamefont {Brukner}},\
  }\bibfield  {title} {\bibinfo {title} {Quantum correlations with no causal
  order},\ }\bibfield  {journal} {\bibinfo  {journal} {Nature Communications}\
  }\textbf {\bibinfo {volume} {3}},\ \href {https://doi.org/10.1038/ncomms2076}
  {10.1038/ncomms2076} (\bibinfo {year} {2012})\BibitemShut {NoStop}%
\bibitem [{\citenamefont {Gu{\'{e} }rin}\ \emph {et~al.}(2016)\citenamefont
  {Gu{\'{e} }rin}, \citenamefont {Feix}, \citenamefont {Ara{\'{u}}jo},\ and\
  \citenamefont {Brukner}}]{Gu_rin_2016}%
  \BibitemOpen
  \bibfield  {author} {\bibinfo {author} {\bibfnamefont {P.~A.}\ \bibnamefont
  {Gu{\'{e} }rin}}, \bibinfo {author} {\bibfnamefont {A.}~\bibnamefont {Feix}},
  \bibinfo {author} {\bibfnamefont {M.}~\bibnamefont {Ara{\'{u}}jo}},\ and\
  \bibinfo {author} {\bibfnamefont {{\v{C}}.}~\bibnamefont {Brukner}},\
  }\bibfield  {title} {\bibinfo {title} {Exponential communication complexity
  advantage from quantum superposition of the direction of communication},\
  }\bibfield  {journal} {\bibinfo  {journal} {Physical Review Letters}\
  }\textbf {\bibinfo {volume} {117}},\ \href
  {https://doi.org/10.1103/physrevlett.117.100502}
  {10.1103/physrevlett.117.100502} (\bibinfo {year} {2016})\BibitemShut
  {NoStop}%
\bibitem [{\citenamefont {Ebler}\ \emph {et~al.}(2018)\citenamefont {Ebler},
  \citenamefont {Salek},\ and\ \citenamefont {Chiribella}}]{ebler2018enhanced}%
  \BibitemOpen
  \bibfield  {author} {\bibinfo {author} {\bibfnamefont {D.}~\bibnamefont
  {Ebler}}, \bibinfo {author} {\bibfnamefont {S.}~\bibnamefont {Salek}},\ and\
  \bibinfo {author} {\bibfnamefont {G.}~\bibnamefont {Chiribella}},\ }\bibfield
   {title} {\bibinfo {title} {Enhanced communication with the assistance of
  indefinite causal order},\ }\href
  {http://dx.doi.org/10.1103/PhysRevLett.120.120502} {\bibfield  {journal}
  {\bibinfo  {journal} {Physical Review Letters}\ }\textbf {\bibinfo {volume}
  {120}},\ \bibinfo {pages} {120502} (\bibinfo {year} {2018})}\BibitemShut
  {NoStop}%
\bibitem [{\citenamefont {Bavaresco}\ \emph {et~al.}(2021)\citenamefont
  {Bavaresco}, \citenamefont {Murao},\ and\ \citenamefont
  {Quintino}}]{Bavaresco2021}%
  \BibitemOpen
  \bibfield  {author} {\bibinfo {author} {\bibfnamefont {J.}~\bibnamefont
  {Bavaresco}}, \bibinfo {author} {\bibfnamefont {M.}~\bibnamefont {Murao}},\
  and\ \bibinfo {author} {\bibfnamefont {M.~T.}\ \bibnamefont {Quintino}},\
  }\bibfield  {title} {\bibinfo {title} {{Strict Hierarchy between Parallel,
  Sequential, and Indefinite-Causal-Order Strategies for Channel
  Discrimination}},\ }\href {https://doi.org/10.1103/PhysRevLett.127.200504}
  {\bibfield  {journal} {\bibinfo  {journal} {Physical Review Letters}\
  }\textbf {\bibinfo {volume} {127}},\ \bibinfo {pages} {200504} (\bibinfo
  {year} {2021})}\BibitemShut {NoStop}%
\bibitem [{\citenamefont {Felce}\ and\ \citenamefont
  {Vedral}(2020)}]{Felce2020}%
  \BibitemOpen
  \bibfield  {author} {\bibinfo {author} {\bibfnamefont {D.}~\bibnamefont
  {Felce}}\ and\ \bibinfo {author} {\bibfnamefont {V.}~\bibnamefont {Vedral}},\
  }\bibfield  {title} {\bibinfo {title} {{Quantum Refrigeration with Indefinite
  Causal Order}},\ }\href {https://doi.org/10.1103/PhysRevLett.125.070603}
  {\bibfield  {journal} {\bibinfo  {journal} {Physical Review Letters}\
  }\textbf {\bibinfo {volume} {125}},\ \bibinfo {pages} {070603} (\bibinfo
  {year} {2020})}\BibitemShut {NoStop}%
\bibitem [{\citenamefont {Quintino}\ \emph {et~al.}(2019)\citenamefont
  {Quintino}, \citenamefont {Dong}, \citenamefont {Shimbo}, \citenamefont
  {Soeda},\ and\ \citenamefont {Murao}}]{Quintino_2019}%
  \BibitemOpen
  \bibfield  {author} {\bibinfo {author} {\bibfnamefont {M.~T.}\ \bibnamefont
  {Quintino}}, \bibinfo {author} {\bibfnamefont {Q.}~\bibnamefont {Dong}},
  \bibinfo {author} {\bibfnamefont {A.}~\bibnamefont {Shimbo}}, \bibinfo
  {author} {\bibfnamefont {A.}~\bibnamefont {Soeda}},\ and\ \bibinfo {author}
  {\bibfnamefont {M.}~\bibnamefont {Murao}},\ }\bibfield  {title} {\bibinfo
  {title} {Reversing unknown quantum transformations: Universal quantum circuit
  for inverting general unitary operations},\ }\bibfield  {journal} {\bibinfo
  {journal} {Physical Review Letters}\ }\textbf {\bibinfo {volume} {123}},\
  \href {https://doi.org/10.1103/physrevlett.123.210502}
  {10.1103/physrevlett.123.210502} (\bibinfo {year} {2019})\BibitemShut
  {NoStop}%
\bibitem [{\citenamefont {Zhao}\ \emph {et~al.}(2020)\citenamefont {Zhao},
  \citenamefont {Yang},\ and\ \citenamefont {Chiribella}}]{Zhao2020a}%
  \BibitemOpen
  \bibfield  {author} {\bibinfo {author} {\bibfnamefont {X.}~\bibnamefont
  {Zhao}}, \bibinfo {author} {\bibfnamefont {Y.}~\bibnamefont {Yang}},\ and\
  \bibinfo {author} {\bibfnamefont {G.}~\bibnamefont {Chiribella}},\ }\bibfield
   {title} {\bibinfo {title} {{Quantum Metrology with Indefinite Causal
  Order}},\ }\href {https://doi.org/10.1103/PhysRevLett.124.190503} {\bibfield
  {journal} {\bibinfo  {journal} {Physical Review Letters}\ }\textbf {\bibinfo
  {volume} {124}},\ \bibinfo {pages} {190503} (\bibinfo {year} {2020})},\
  \Eprint {https://arxiv.org/abs/1912.02449} {arXiv:1912.02449} \BibitemShut
  {NoStop}%
\bibitem [{\citenamefont {Gottesman}(2009)}]{gottesman2009introduction}%
  \BibitemOpen
  \bibfield  {author} {\bibinfo {author} {\bibfnamefont {D.}~\bibnamefont
  {Gottesman}},\ }\href@noop {} {\bibinfo {title} {An introduction to quantum
  error correction and fault-tolerant quantum computation}} (\bibinfo {year}
  {2009}),\ \Eprint {https://arxiv.org/abs/0904.2557} {arXiv:0904.2557
  [quant-ph]} \BibitemShut {NoStop}%
\bibitem [{\citenamefont {Gottesman}\ and\ \citenamefont
  {Chuang}(1999)}]{Gottesman_1999}%
  \BibitemOpen
  \bibfield  {author} {\bibinfo {author} {\bibfnamefont {D.}~\bibnamefont
  {Gottesman}}\ and\ \bibinfo {author} {\bibfnamefont {I.~L.}\ \bibnamefont
  {Chuang}},\ }\bibfield  {title} {\bibinfo {title} {Demonstrating the
  viability of universal quantum computation using teleportation and
  single-qubit operations},\ }\href {https://doi.org/10.1038/46503} {\bibfield
  {journal} {\bibinfo  {journal} {Nature}\ }\textbf {\bibinfo {volume} {402}},\
  \bibinfo {pages} {390} (\bibinfo {year} {1999})}\BibitemShut {NoStop}%
\bibitem [{\citenamefont {Zhou}\ \emph {et~al.}(2000)\citenamefont {Zhou},
  \citenamefont {Leung},\ and\ \citenamefont {Chuang}}]{Zhou_2000}%
  \BibitemOpen
  \bibfield  {author} {\bibinfo {author} {\bibfnamefont {X.}~\bibnamefont
  {Zhou}}, \bibinfo {author} {\bibfnamefont {D.~W.}\ \bibnamefont {Leung}},\
  and\ \bibinfo {author} {\bibfnamefont {I.~L.}\ \bibnamefont {Chuang}},\
  }\bibfield  {title} {\bibinfo {title} {Methodology for quantum logic gate
  construction},\ }\bibfield  {journal} {\bibinfo  {journal} {Physical Review
  A}\ }\textbf {\bibinfo {volume} {62}},\ \href
  {https://doi.org/10.1103/physreva.62.052316} {10.1103/physreva.62.052316}
  (\bibinfo {year} {2000})\BibitemShut {NoStop}%
\bibitem [{\citenamefont {Aaronson}\ and\ \citenamefont
  {Gottesman}(2004)}]{Aaronson_2004}%
  \BibitemOpen
  \bibfield  {author} {\bibinfo {author} {\bibfnamefont {S.}~\bibnamefont
  {Aaronson}}\ and\ \bibinfo {author} {\bibfnamefont {D.}~\bibnamefont
  {Gottesman}},\ }\bibfield  {title} {\bibinfo {title} {Improved simulation of
  stabilizer circuits},\ }\bibfield  {journal} {\bibinfo  {journal} {Physical
  Review A}\ }\textbf {\bibinfo {volume} {70}},\ \href
  {https://doi.org/10.1103/physreva.70.052328} {10.1103/physreva.70.052328}
  (\bibinfo {year} {2004})\BibitemShut {NoStop}%
\bibitem [{\citenamefont {Bravyi}\ \emph {et~al.}(2016)\citenamefont {Bravyi},
  \citenamefont {Smith},\ and\ \citenamefont {Smolin}}]{Bravyi2016}%
  \BibitemOpen
  \bibfield  {author} {\bibinfo {author} {\bibfnamefont {S.}~\bibnamefont
  {Bravyi}}, \bibinfo {author} {\bibfnamefont {G.}~\bibnamefont {Smith}},\ and\
  \bibinfo {author} {\bibfnamefont {J.~A.}\ \bibnamefont {Smolin}},\ }\bibfield
   {title} {\bibinfo {title} {{Trading classical and quantum computational
  resources}},\ }\href {https://doi.org/10.1103/PhysRevX.6.021043} {\bibfield
  {journal} {\bibinfo  {journal} {Physical Review X}\ }\textbf {\bibinfo
  {volume} {6}},\ \bibinfo {pages} {1} (\bibinfo {year} {2016})},\ \Eprint
  {https://arxiv.org/abs/1506.01396} {arXiv:1506.01396} \BibitemShut {NoStop}%
\bibitem [{\citenamefont {Bravyi}\ \emph {et~al.}(2019)\citenamefont {Bravyi},
  \citenamefont {Browne}, \citenamefont {Calpin}, \citenamefont {Campbell},
  \citenamefont {Gosset},\ and\ \citenamefont {Howard}}]{Bravyi_2019}%
  \BibitemOpen
  \bibfield  {author} {\bibinfo {author} {\bibfnamefont {S.}~\bibnamefont
  {Bravyi}}, \bibinfo {author} {\bibfnamefont {D.}~\bibnamefont {Browne}},
  \bibinfo {author} {\bibfnamefont {P.}~\bibnamefont {Calpin}}, \bibinfo
  {author} {\bibfnamefont {E.}~\bibnamefont {Campbell}}, \bibinfo {author}
  {\bibfnamefont {D.}~\bibnamefont {Gosset}},\ and\ \bibinfo {author}
  {\bibfnamefont {M.}~\bibnamefont {Howard}},\ }\bibfield  {title} {\bibinfo
  {title} {Simulation of quantum circuits by low-rank stabilizer
  decompositions},\ }\href {https://doi.org/10.22331/q-2019-09-02-181}
  {\bibfield  {journal} {\bibinfo  {journal} {Quantum}\ }\textbf {\bibinfo
  {volume} {3}},\ \bibinfo {pages} {181} (\bibinfo {year} {2019})}\BibitemShut
  {NoStop}%
\bibitem [{\citenamefont {Howard}\ and\ \citenamefont
  {Campbell}(2017{\natexlab{a}})}]{Howard_2017}%
  \BibitemOpen
  \bibfield  {author} {\bibinfo {author} {\bibfnamefont {M.}~\bibnamefont
  {Howard}}\ and\ \bibinfo {author} {\bibfnamefont {E.}~\bibnamefont
  {Campbell}},\ }\bibfield  {title} {\bibinfo {title} {{Application of a
  Resource Theory for Magic States to Fault-Tolerant Quantum Computing}},\
  }\bibfield  {journal} {\bibinfo  {journal} {Physical Review Letters}\
  }\textbf {\bibinfo {volume} {118}},\ \href
  {https://doi.org/10.1103/PhysRevLett.118.090501}
  {10.1103/PhysRevLett.118.090501} (\bibinfo {year} {2017}{\natexlab{a}}),\
  \Eprint {https://arxiv.org/abs/1609.07488} {arXiv:1609.07488} \BibitemShut
  {NoStop}%
\bibitem [{\citenamefont {Chitambar}\ and\ \citenamefont
  {Gour}(2019)}]{RMP_Chitambar_2019}%
  \BibitemOpen
  \bibfield  {author} {\bibinfo {author} {\bibfnamefont {E.}~\bibnamefont
  {Chitambar}}\ and\ \bibinfo {author} {\bibfnamefont {G.}~\bibnamefont
  {Gour}},\ }\bibfield  {title} {\bibinfo {title} {Quantum resource theories},\
  }\bibfield  {journal} {\bibinfo  {journal} {Reviews of Modern Physics}\
  }\textbf {\bibinfo {volume} {91}},\ \href
  {https://doi.org/10.1103/revmodphys.91.025001} {10.1103/revmodphys.91.025001}
  (\bibinfo {year} {2019})\BibitemShut {NoStop}%
\bibitem [{\citenamefont {Rubino}\ \emph {et~al.}(2017)\citenamefont {Rubino},
  \citenamefont {Rozema}, \citenamefont {Feix}, \citenamefont {Ara{\'{u}}jo},
  \citenamefont {Zeuner}, \citenamefont {Procopio}, \citenamefont {Brukner},\
  and\ \citenamefont {Walther}}]{Rubino2017}%
  \BibitemOpen
  \bibfield  {author} {\bibinfo {author} {\bibfnamefont {G.}~\bibnamefont
  {Rubino}}, \bibinfo {author} {\bibfnamefont {L.~A.}\ \bibnamefont {Rozema}},
  \bibinfo {author} {\bibfnamefont {A.}~\bibnamefont {Feix}}, \bibinfo {author}
  {\bibfnamefont {M.}~\bibnamefont {Ara{\'{u}}jo}}, \bibinfo {author}
  {\bibfnamefont {J.~M.}\ \bibnamefont {Zeuner}}, \bibinfo {author}
  {\bibfnamefont {L.~M.}\ \bibnamefont {Procopio}}, \bibinfo {author}
  {\bibfnamefont {{\v{C}}.}~\bibnamefont {Brukner}},\ and\ \bibinfo {author}
  {\bibfnamefont {P.}~\bibnamefont {Walther}},\ }\bibfield  {title} {\bibinfo
  {title} {{Experimental verification of an indefinite causal order}},\ }\href
  {https://doi.org/10.1126/sciadv.1602589} {\bibfield  {journal} {\bibinfo
  {journal} {Science Advances}\ }\textbf {\bibinfo {volume} {3}},\ \bibinfo
  {pages} {e1602589} (\bibinfo {year} {2017})}\BibitemShut {NoStop}%
\bibitem [{\citenamefont {Yin}\ \emph {et~al.}(2023)\citenamefont {Yin},
  \citenamefont {Zhao}, \citenamefont {Yang}, \citenamefont {Guo},
  \citenamefont {Zhang}, \citenamefont {Li}, \citenamefont {Han}, \citenamefont
  {Liu}, \citenamefont {Xu}, \citenamefont {Chiribella} \emph
  {et~al.}}]{Yin2023experimental}%
  \BibitemOpen
  \bibfield  {author} {\bibinfo {author} {\bibfnamefont {P.}~\bibnamefont
  {Yin}}, \bibinfo {author} {\bibfnamefont {X.}~\bibnamefont {Zhao}}, \bibinfo
  {author} {\bibfnamefont {Y.}~\bibnamefont {Yang}}, \bibinfo {author}
  {\bibfnamefont {Y.}~\bibnamefont {Guo}}, \bibinfo {author} {\bibfnamefont
  {W.-H.}\ \bibnamefont {Zhang}}, \bibinfo {author} {\bibfnamefont {G.-C.}\
  \bibnamefont {Li}}, \bibinfo {author} {\bibfnamefont {Y.-J.}\ \bibnamefont
  {Han}}, \bibinfo {author} {\bibfnamefont {B.-H.}\ \bibnamefont {Liu}},
  \bibinfo {author} {\bibfnamefont {J.-S.}\ \bibnamefont {Xu}}, \bibinfo
  {author} {\bibfnamefont {G.}~\bibnamefont {Chiribella}}, \emph {et~al.},\
  }\bibfield  {title} {\bibinfo {title} {Experimental super-heisenberg quantum
  metrology with indefinite gate order},\ }\href
  {https://www.nature.com/articles/s41567-023-02046-y} {\bibfield  {journal}
  {\bibinfo  {journal} {Nature Physics}\ }\textbf {\bibinfo {volume} {19}},\
  \bibinfo {pages} {1122} (\bibinfo {year} {2023})}\BibitemShut {NoStop}%
\bibitem [{\citenamefont {Guo}\ \emph {et~al.}(2020)\citenamefont {Guo},
  \citenamefont {Hu}, \citenamefont {Hou}, \citenamefont {Cao}, \citenamefont
  {Cui}, \citenamefont {Liu}, \citenamefont {Huang}, \citenamefont {Li},
  \citenamefont {Guo},\ and\ \citenamefont {Chiribella}}]{Guo2020experimental}%
  \BibitemOpen
  \bibfield  {author} {\bibinfo {author} {\bibfnamefont {Y.}~\bibnamefont
  {Guo}}, \bibinfo {author} {\bibfnamefont {X.-M.}\ \bibnamefont {Hu}},
  \bibinfo {author} {\bibfnamefont {Z.-B.}\ \bibnamefont {Hou}}, \bibinfo
  {author} {\bibfnamefont {H.}~\bibnamefont {Cao}}, \bibinfo {author}
  {\bibfnamefont {J.-M.}\ \bibnamefont {Cui}}, \bibinfo {author} {\bibfnamefont
  {B.-H.}\ \bibnamefont {Liu}}, \bibinfo {author} {\bibfnamefont {Y.-F.}\
  \bibnamefont {Huang}}, \bibinfo {author} {\bibfnamefont {C.-F.}\ \bibnamefont
  {Li}}, \bibinfo {author} {\bibfnamefont {G.-C.}\ \bibnamefont {Guo}},\ and\
  \bibinfo {author} {\bibfnamefont {G.}~\bibnamefont {Chiribella}},\ }\bibfield
   {title} {\bibinfo {title} {Experimental transmission of quantum information
  using a superposition of causal orders},\ }\href
  {https://doi.org/10.1103/PhysRevLett.124.030502} {\bibfield  {journal}
  {\bibinfo  {journal} {Physical Review Letter}\ }\textbf {\bibinfo {volume}
  {124}},\ \bibinfo {pages} {030502} (\bibinfo {year} {2020})}\BibitemShut
  {NoStop}%
\bibitem [{\citenamefont {Nie}\ \emph {et~al.}(2022)\citenamefont {Nie},
  \citenamefont {Zhu}, \citenamefont {Huang}, \citenamefont {Tang},
  \citenamefont {Long}, \citenamefont {Lin}, \citenamefont {Tian},
  \citenamefont {Qiu}, \citenamefont {Xi}, \citenamefont {Yang}, \citenamefont
  {Li}, \citenamefont {Dong}, \citenamefont {Xin},\ and\ \citenamefont
  {Lu}}]{Nie2022}%
  \BibitemOpen
  \bibfield  {author} {\bibinfo {author} {\bibfnamefont {X.}~\bibnamefont
  {Nie}}, \bibinfo {author} {\bibfnamefont {X.}~\bibnamefont {Zhu}}, \bibinfo
  {author} {\bibfnamefont {K.}~\bibnamefont {Huang}}, \bibinfo {author}
  {\bibfnamefont {K.}~\bibnamefont {Tang}}, \bibinfo {author} {\bibfnamefont
  {X.}~\bibnamefont {Long}}, \bibinfo {author} {\bibfnamefont {Z.}~\bibnamefont
  {Lin}}, \bibinfo {author} {\bibfnamefont {Y.}~\bibnamefont {Tian}}, \bibinfo
  {author} {\bibfnamefont {C.}~\bibnamefont {Qiu}}, \bibinfo {author}
  {\bibfnamefont {C.}~\bibnamefont {Xi}}, \bibinfo {author} {\bibfnamefont
  {X.}~\bibnamefont {Yang}}, \bibinfo {author} {\bibfnamefont {J.}~\bibnamefont
  {Li}}, \bibinfo {author} {\bibfnamefont {Y.}~\bibnamefont {Dong}}, \bibinfo
  {author} {\bibfnamefont {T.}~\bibnamefont {Xin}},\ and\ \bibinfo {author}
  {\bibfnamefont {D.}~\bibnamefont {Lu}},\ }\bibfield  {title} {\bibinfo
  {title} {{Experimental Realization of a Quantum Refrigerator Driven by
  Indefinite Causal Orders}},\ }\href
  {https://doi.org/10.1103/PhysRevLett.129.100603} {\bibfield  {journal}
  {\bibinfo  {journal} {Physical Review Letters}\ }\textbf {\bibinfo {volume}
  {129}},\ \bibinfo {pages} {100603} (\bibinfo {year} {2022})}\BibitemShut
  {NoStop}%
\bibitem [{\citenamefont {Goswami}\ \emph {et~al.}(2018)\citenamefont
  {Goswami}, \citenamefont {Giarmatzi}, \citenamefont {Kewming}, \citenamefont
  {Costa}, \citenamefont {Branciard}, \citenamefont {Romero},\ and\
  \citenamefont {White}}]{Goswami2018}%
  \BibitemOpen
  \bibfield  {author} {\bibinfo {author} {\bibfnamefont {K.}~\bibnamefont
  {Goswami}}, \bibinfo {author} {\bibfnamefont {C.}~\bibnamefont {Giarmatzi}},
  \bibinfo {author} {\bibfnamefont {M.}~\bibnamefont {Kewming}}, \bibinfo
  {author} {\bibfnamefont {F.}~\bibnamefont {Costa}}, \bibinfo {author}
  {\bibfnamefont {C.}~\bibnamefont {Branciard}}, \bibinfo {author}
  {\bibfnamefont {J.}~\bibnamefont {Romero}},\ and\ \bibinfo {author}
  {\bibfnamefont {A.~G.}\ \bibnamefont {White}},\ }\bibfield  {title} {\bibinfo
  {title} {{Indefinite Causal Order in a Quantum Switch}},\ }\href
  {https://doi.org/10.1103/PhysRevLett.121.090503} {\bibfield  {journal}
  {\bibinfo  {journal} {Physical Review Letters}\ }\textbf {\bibinfo {volume}
  {121}},\ \bibinfo {pages} {090503} (\bibinfo {year} {2018})}\BibitemShut
  {NoStop}%
\bibitem [{\citenamefont {Veitch}\ \emph {et~al.}(2014)\citenamefont {Veitch},
  \citenamefont {{Hamed Mousavian}}, \citenamefont {Gottesman},\ and\
  \citenamefont {Emerson}}]{Veitch2014}%
  \BibitemOpen
  \bibfield  {author} {\bibinfo {author} {\bibfnamefont {V.}~\bibnamefont
  {Veitch}}, \bibinfo {author} {\bibfnamefont {S.~A.}\ \bibnamefont {{Hamed
  Mousavian}}}, \bibinfo {author} {\bibfnamefont {D.}~\bibnamefont
  {Gottesman}},\ and\ \bibinfo {author} {\bibfnamefont {J.}~\bibnamefont
  {Emerson}},\ }\bibfield  {title} {\bibinfo {title} {{The resource theory of
  stabilizer quantum computation}},\ }\href
  {https://doi.org/10.1088/1367-2630/16/1/013009} {\bibfield  {journal}
  {\bibinfo  {journal} {New Journal of Physics}\ }\textbf {\bibinfo {volume}
  {16}},\ \bibinfo {pages} {1} (\bibinfo {year} {2014})},\ \Eprint
  {https://arxiv.org/abs/arXiv:1307.7171v1} {arXiv:arXiv:1307.7171v1}
  \BibitemShut {NoStop}%
\bibitem [{\citenamefont {Howard}\ and\ \citenamefont
  {Campbell}(2017{\natexlab{b}})}]{Howard_2017}%
  \BibitemOpen
  \bibfield  {author} {\bibinfo {author} {\bibfnamefont {M.}~\bibnamefont
  {Howard}}\ and\ \bibinfo {author} {\bibfnamefont {E.}~\bibnamefont
  {Campbell}},\ }\bibfield  {title} {\bibinfo {title} {Application of a
  resource theory for magic states to fault-tolerant quantum computing},\
  }\bibfield  {journal} {\bibinfo  {journal} {Physical Review Letters}\
  }\textbf {\bibinfo {volume} {118}},\ \href
  {https://doi.org/10.1103/physrevlett.118.090501}
  {10.1103/physrevlett.118.090501} (\bibinfo {year}
  {2017}{\natexlab{b}})\BibitemShut {NoStop}%
\bibitem [{\citenamefont {Seddon}\ and\ \citenamefont
  {Campbell}(2019)}]{Seddon_2019}%
  \BibitemOpen
  \bibfield  {author} {\bibinfo {author} {\bibfnamefont {J.~R.}\ \bibnamefont
  {Seddon}}\ and\ \bibinfo {author} {\bibfnamefont {E.~T.}\ \bibnamefont
  {Campbell}},\ }\bibfield  {title} {\bibinfo {title} {Quantifying magic for
  multi-qubit operations},\ }\href {https://doi.org/10.1098/rspa.2019.0251}
  {\bibfield  {journal} {\bibinfo  {journal} {Proceedings of the Royal Society
  A: Mathematical, Physical and Engineering Sciences}\ }\textbf {\bibinfo
  {volume} {475}},\ \bibinfo {pages} {20190251} (\bibinfo {year}
  {2019})}\BibitemShut {NoStop}%
\bibitem [{\citenamefont {Heinrich}\ and\ \citenamefont
  {Gross}(2019)}]{Heinrich_2019}%
  \BibitemOpen
  \bibfield  {author} {\bibinfo {author} {\bibfnamefont {M.}~\bibnamefont
  {Heinrich}}\ and\ \bibinfo {author} {\bibfnamefont {D.}~\bibnamefont
  {Gross}},\ }\bibfield  {title} {\bibinfo {title} {Robustness of magic and
  symmetries of the stabiliser polytope},\ }\href
  {https://doi.org/10.22331/q-2019-04-08-132} {\bibfield  {journal} {\bibinfo
  {journal} {Quantum}\ }\textbf {\bibinfo {volume} {3}},\ \bibinfo {pages}
  {132} (\bibinfo {year} {2019})}\BibitemShut {NoStop}%
\bibitem [{\citenamefont {Wang}\ \emph {et~al.}(2018)\citenamefont {Wang},
  \citenamefont {Wilde},\ and\ \citenamefont {Su}}]{Wang2018}%
  \BibitemOpen
  \bibfield  {author} {\bibinfo {author} {\bibfnamefont {X.}~\bibnamefont
  {Wang}}, \bibinfo {author} {\bibfnamefont {M.~M.}\ \bibnamefont {Wilde}},\
  and\ \bibinfo {author} {\bibfnamefont {Y.}~\bibnamefont {Su}},\ }\bibfield
  {title} {\bibinfo {title} {{Efficiently computable bounds for magic state
  distillation}},\ }\href {https://doi.org/10.1103/PhysRevLett.124.090505}
  {\bibfield  {journal} {\bibinfo  {journal} {Physical Review Letters}\
  }\textbf {\bibinfo {volume} {124}},\ \bibinfo {pages} {090505} (\bibinfo
  {year} {2018})},\ \Eprint {https://arxiv.org/abs/1812.10145}
  {arXiv:1812.10145} \BibitemShut {NoStop}%
\bibitem [{\citenamefont {Saxena}\ and\ \citenamefont
  {Gour}(2022)}]{Saxena2022}%
  \BibitemOpen
  \bibfield  {author} {\bibinfo {author} {\bibfnamefont {G.}~\bibnamefont
  {Saxena}}\ and\ \bibinfo {author} {\bibfnamefont {G.}~\bibnamefont {Gour}},\
  }\bibfield  {title} {\bibinfo {title} {{Quantifying multiqubit magic channels
  with completely stabilizer-preserving operations}},\ }\href
  {https://doi.org/10.1103/PhysRevA.106.042422} {\bibfield  {journal} {\bibinfo
   {journal} {Physical Review A}\ }\textbf {\bibinfo {volume} {106}},\ \bibinfo
  {pages} {042422} (\bibinfo {year} {2022})},\ \Eprint
  {https://arxiv.org/abs/2202.07867} {arXiv:2202.07867} \BibitemShut {NoStop}%
\bibitem [{\citenamefont {Beverland}\ \emph {et~al.}(2020)\citenamefont
  {Beverland}, \citenamefont {Campbell}, \citenamefont {Howard},\ and\
  \citenamefont {Kliuchnikov}}]{Beverland2019}%
  \BibitemOpen
  \bibfield  {author} {\bibinfo {author} {\bibfnamefont {M.}~\bibnamefont
  {Beverland}}, \bibinfo {author} {\bibfnamefont {E.}~\bibnamefont {Campbell}},
  \bibinfo {author} {\bibfnamefont {M.}~\bibnamefont {Howard}},\ and\ \bibinfo
  {author} {\bibfnamefont {V.}~\bibnamefont {Kliuchnikov}},\ }\bibfield
  {title} {\bibinfo {title} {{Lower bounds on the non-Clifford resources for
  quantum computations}},\ }\href {https://doi.org/10.1088/2058-9565/ab8963}
  {\bibfield  {journal} {\bibinfo  {journal} {Quantum Science and Technology}\
  }\textbf {\bibinfo {volume} {5}},\ \bibinfo {pages} {035009} (\bibinfo {year}
  {2020})},\ \Eprint {https://arxiv.org/abs/1904.01124} {arXiv:1904.01124}
  \BibitemShut {NoStop}%
\bibitem [{\citenamefont {Jiang}\ and\ \citenamefont {Wang}(2023)}]{Jiang2021}%
  \BibitemOpen
  \bibfield  {author} {\bibinfo {author} {\bibfnamefont {J.}~\bibnamefont
  {Jiang}}\ and\ \bibinfo {author} {\bibfnamefont {X.}~\bibnamefont {Wang}},\
  }\bibfield  {title} {\bibinfo {title} {{Lower Bound for the T Count Via
  Unitary Stabilizer Nullity}},\ }\href
  {https://doi.org/10.1103/PhysRevApplied.19.034052} {\bibfield  {journal}
  {\bibinfo  {journal} {Physical Review Applied}\ }\textbf {\bibinfo {volume}
  {19}},\ \bibinfo {pages} {034052} (\bibinfo {year} {2023})},\ \Eprint
  {https://arxiv.org/abs/2103.09999} {arXiv:2103.09999} \BibitemShut {NoStop}%
\bibitem [{\citenamefont {Seddon}\ \emph {et~al.}(2021)\citenamefont {Seddon},
  \citenamefont {Regula}, \citenamefont {Pashayan}, \citenamefont {Ouyang},\
  and\ \citenamefont {Campbell}}]{Seddon2021}%
  \BibitemOpen
  \bibfield  {author} {\bibinfo {author} {\bibfnamefont {J.~R.}\ \bibnamefont
  {Seddon}}, \bibinfo {author} {\bibfnamefont {B.}~\bibnamefont {Regula}},
  \bibinfo {author} {\bibfnamefont {H.}~\bibnamefont {Pashayan}}, \bibinfo
  {author} {\bibfnamefont {Y.}~\bibnamefont {Ouyang}},\ and\ \bibinfo {author}
  {\bibfnamefont {E.~T.}\ \bibnamefont {Campbell}},\ }\bibfield  {title}
  {\bibinfo {title} {{Quantifying Quantum Speedups: Improved Classical
  Simulation From Tighter Magic Monotones}},\ }\href
  {https://doi.org/10.1103/PRXQuantum.2.010345} {\bibfield  {journal} {\bibinfo
   {journal} {PRX Quantum}\ }\textbf {\bibinfo {volume} {2}},\ \bibinfo {pages}
  {010345} (\bibinfo {year} {2021})},\ \Eprint
  {https://arxiv.org/abs/2002.06181} {arXiv:2002.06181} \BibitemShut {NoStop}%
\bibitem [{\citenamefont {Wang}\ \emph {et~al.}(2019)\citenamefont {Wang},
  \citenamefont {Wilde},\ and\ \citenamefont {Su}}]{WWS19}%
  \BibitemOpen
  \bibfield  {author} {\bibinfo {author} {\bibfnamefont {X.}~\bibnamefont
  {Wang}}, \bibinfo {author} {\bibfnamefont {M.~M.}\ \bibnamefont {Wilde}},\
  and\ \bibinfo {author} {\bibfnamefont {Y.}~\bibnamefont {Su}},\ }\bibfield
  {title} {\bibinfo {title} {{Quantifying the magic of quantum channels}},\
  }\href {https://doi.org/10.1088/1367-2630/ab451d} {\bibfield  {journal}
  {\bibinfo  {journal} {New Journal of Physics}\ }\textbf {\bibinfo {volume}
  {21}},\ \bibinfo {pages} {103002} (\bibinfo {year} {2019})},\ \Eprint
  {https://arxiv.org/abs/1903.04483} {arXiv:1903.04483} \BibitemShut {NoStop}%
\bibitem [{\citenamefont {Haug}\ and\ \citenamefont {Kim}(2023)}]{Haug2023}%
  \BibitemOpen
  \bibfield  {author} {\bibinfo {author} {\bibfnamefont {T.}~\bibnamefont
  {Haug}}\ and\ \bibinfo {author} {\bibfnamefont {M.}~\bibnamefont {Kim}},\
  }\bibfield  {title} {\bibinfo {title} {{Scalable Measures of Magic Resource
  for Quantum Computers}},\ }\href
  {https://doi.org/10.1103/PRXQuantum.4.010301} {\bibfield  {journal} {\bibinfo
   {journal} {PRX Quantum}\ }\textbf {\bibinfo {volume} {4}},\ \bibinfo {pages}
  {010301} (\bibinfo {year} {2023})}\BibitemShut {NoStop}%
\bibitem [{\citenamefont {Leone}\ \emph {et~al.}(2022)\citenamefont {Leone},
  \citenamefont {Oliviero},\ and\ \citenamefont {Hamma}}]{Leone2022}%
  \BibitemOpen
  \bibfield  {author} {\bibinfo {author} {\bibfnamefont {L.}~\bibnamefont
  {Leone}}, \bibinfo {author} {\bibfnamefont {S.~F.~E.}\ \bibnamefont
  {Oliviero}},\ and\ \bibinfo {author} {\bibfnamefont {A.}~\bibnamefont
  {Hamma}},\ }\bibfield  {title} {\bibinfo {title} {{Stabilizer r{\'{e}}nyi
  entropy}},\ }\href {http://dx.doi.org/10.1103/PhysRevLett.128.050402}
  {\bibfield  {journal} {\bibinfo  {journal} {Physical Review Letters}\
  }\textbf {\bibinfo {volume} {128}},\ \bibinfo {pages} {50402} (\bibinfo
  {year} {2022})}\BibitemShut {NoStop}%
\bibitem [{\citenamefont {Oliviero}\ \emph {et~al.}(2022)\citenamefont
  {Oliviero}, \citenamefont {Leone}, \citenamefont {Hamma},\ and\ \citenamefont
  {Lloyd}}]{Oliviero2022}%
  \BibitemOpen
  \bibfield  {author} {\bibinfo {author} {\bibfnamefont {S.~F.~E.}\
  \bibnamefont {Oliviero}}, \bibinfo {author} {\bibfnamefont {L.}~\bibnamefont
  {Leone}}, \bibinfo {author} {\bibfnamefont {A.}~\bibnamefont {Hamma}},\ and\
  \bibinfo {author} {\bibfnamefont {S.}~\bibnamefont {Lloyd}},\ }\bibfield
  {title} {\bibinfo {title} {{Measuring magic on a quantum processor}},\ }\href
  {http://dx.doi.org/10.1038/s41534-022-00666-5} {\bibfield  {journal}
  {\bibinfo  {journal} {npj Quantum Information}\ }\textbf {\bibinfo {volume}
  {8}},\ \bibinfo {pages} {148} (\bibinfo {year} {2022})}\BibitemShut {NoStop}%
\bibitem [{\citenamefont {Liu}\ and\ \citenamefont {Winter}(2022)}]{Liu2022d}%
  \BibitemOpen
  \bibfield  {author} {\bibinfo {author} {\bibfnamefont {Z.-W.}\ \bibnamefont
  {Liu}}\ and\ \bibinfo {author} {\bibfnamefont {A.}~\bibnamefont {Winter}},\
  }\bibfield  {title} {\bibinfo {title} {{Many-Body Quantum Magic}},\ }\href
  {https://doi.org/10.1103/PRXQuantum.3.020333} {\bibfield  {journal} {\bibinfo
   {journal} {PRX Quantum}\ }\textbf {\bibinfo {volume} {3}},\ \bibinfo {pages}
  {020333} (\bibinfo {year} {2022})}\BibitemShut {NoStop}%
\bibitem [{\citenamefont {Bu}\ \emph {et~al.}(2023)\citenamefont {Bu},
  \citenamefont {Gu},\ and\ \citenamefont {Jaffe}}]{Bu2023a}%
  \BibitemOpen
  \bibfield  {author} {\bibinfo {author} {\bibfnamefont {K.}~\bibnamefont
  {Bu}}, \bibinfo {author} {\bibfnamefont {W.}~\bibnamefont {Gu}},\ and\
  \bibinfo {author} {\bibfnamefont {A.}~\bibnamefont {Jaffe}},\ }\bibfield
  {title} {\bibinfo {title} {{Quantum entropy and central limit theorem}},\
  }\href {https://doi.org/10.1073/pnas.2304589120} {\bibfield  {journal}
  {\bibinfo  {journal} {Proceedings of the National Academy of Sciences}\
  }\textbf {\bibinfo {volume} {120}},\ \bibinfo {pages} {e2304589120} (\bibinfo
  {year} {2023})}\BibitemShut {NoStop}%
\bibitem [{\citenamefont {Haug}\ and\ \citenamefont
  {Piroli}(2023)}]{Haug2023b}%
  \BibitemOpen
  \bibfield  {author} {\bibinfo {author} {\bibfnamefont {T.}~\bibnamefont
  {Haug}}\ and\ \bibinfo {author} {\bibfnamefont {L.}~\bibnamefont {Piroli}},\
  }\bibfield  {title} {\bibinfo {title} {{Quantifying nonstabilizerness of
  matrix product states}},\ }\href
  {https://journals.aps.org/prb/abstract/10.1103/PhysRevB.107.035148}
  {\bibfield  {journal} {\bibinfo  {journal} {Physical Review B}\ }\textbf
  {\bibinfo {volume} {107}},\ \bibinfo {pages} {35148} (\bibinfo {year}
  {2023})}\BibitemShut {NoStop}%
\bibitem [{\citenamefont {Lami}\ and\ \citenamefont
  {Collura}(2023)}]{Lami2023}%
  \BibitemOpen
  \bibfield  {author} {\bibinfo {author} {\bibfnamefont {G.}~\bibnamefont
  {Lami}}\ and\ \bibinfo {author} {\bibfnamefont {M.}~\bibnamefont {Collura}},\
  }\bibfield  {title} {\bibinfo {title} {{Nonstabilizerness via Perfect Pauli
  Sampling of Matrix Product States}},\ }\href
  {https://journals.aps.org/prl/abstract/10.1103/PhysRevLett.131.180401}
  {\bibfield  {journal} {\bibinfo  {journal} {Physical Review Letters}\
  }\textbf {\bibinfo {volume} {131}},\ \bibinfo {pages} {180401} (\bibinfo
  {year} {2023})}\BibitemShut {NoStop}%
\bibitem [{\citenamefont {Ahmadi}\ \emph {et~al.}(2018)\citenamefont {Ahmadi},
  \citenamefont {Dang}, \citenamefont {Gour},\ and\ \citenamefont
  {Sanders}}]{Ahmadi_2018}%
  \BibitemOpen
  \bibfield  {author} {\bibinfo {author} {\bibfnamefont {M.}~\bibnamefont
  {Ahmadi}}, \bibinfo {author} {\bibfnamefont {H.~B.}\ \bibnamefont {Dang}},
  \bibinfo {author} {\bibfnamefont {G.}~\bibnamefont {Gour}},\ and\ \bibinfo
  {author} {\bibfnamefont {B.~C.}\ \bibnamefont {Sanders}},\ }\bibfield
  {title} {\bibinfo {title} {Quantification and manipulation of magic states},\
  }\bibfield  {journal} {\bibinfo  {journal} {Physical Review A}\ }\textbf
  {\bibinfo {volume} {97}},\ \href {https://doi.org/10.1103/physreva.97.062332}
  {10.1103/physreva.97.062332} (\bibinfo {year} {2018})\BibitemShut {NoStop}%
\bibitem [{\citenamefont {Leifer}\ \emph {et~al.}(2003)\citenamefont {Leifer},
  \citenamefont {Henderson},\ and\ \citenamefont {Linden}}]{Leifer_2003}%
  \BibitemOpen
  \bibfield  {author} {\bibinfo {author} {\bibfnamefont {M.~S.}\ \bibnamefont
  {Leifer}}, \bibinfo {author} {\bibfnamefont {L.}~\bibnamefont {Henderson}},\
  and\ \bibinfo {author} {\bibfnamefont {N.}~\bibnamefont {Linden}},\
  }\bibfield  {title} {\bibinfo {title} {Optimal entanglement generation from
  quantum operations},\ }\bibfield  {journal} {\bibinfo  {journal} {Physical
  Review A}\ }\textbf {\bibinfo {volume} {67}},\ \href
  {https://doi.org/10.1103/physreva.67.012306} {10.1103/physreva.67.012306}
  (\bibinfo {year} {2003})\BibitemShut {NoStop}%
\bibitem [{\citenamefont {Bennett}\ \emph {et~al.}(2003)\citenamefont
  {Bennett}, \citenamefont {Harrow}, \citenamefont {Leung},\ and\ \citenamefont
  {Smolin}}]{bennett2003capacities}%
  \BibitemOpen
  \bibfield  {author} {\bibinfo {author} {\bibfnamefont {C.~H.}\ \bibnamefont
  {Bennett}}, \bibinfo {author} {\bibfnamefont {A.~W.}\ \bibnamefont {Harrow}},
  \bibinfo {author} {\bibfnamefont {D.~W.}\ \bibnamefont {Leung}},\ and\
  \bibinfo {author} {\bibfnamefont {J.~A.}\ \bibnamefont {Smolin}},\ }\bibfield
   {title} {\bibinfo {title} {On the capacities of bipartite hamiltonians and
  unitary gates},\ }\href
  {https://ieeexplore.ieee.org/abstract/document/1214070} {\bibfield  {journal}
  {\bibinfo  {journal} {IEEE Transactions on Information Theory}\ }\textbf
  {\bibinfo {volume} {49}},\ \bibinfo {pages} {1895} (\bibinfo {year}
  {2003})}\BibitemShut {NoStop}%
\bibitem [{\citenamefont {Stahlke}(2014)}]{Stahlke_2014}%
  \BibitemOpen
  \bibfield  {author} {\bibinfo {author} {\bibfnamefont {D.}~\bibnamefont
  {Stahlke}},\ }\bibfield  {title} {\bibinfo {title} {Quantum interference as a
  resource for quantum speedup},\ }\bibfield  {journal} {\bibinfo  {journal}
  {Physical Review A}\ }\textbf {\bibinfo {volume} {90}},\ \href
  {https://doi.org/10.1103/physreva.90.022302} {10.1103/physreva.90.022302}
  (\bibinfo {year} {2014})\BibitemShut {NoStop}%
\bibitem [{\citenamefont {Kukulski}\ \emph {et~al.}(2021)\citenamefont
  {Kukulski}, \citenamefont {Nechita}, \citenamefont {Pawela}, \citenamefont
  {Puchala},\ and\ \citenamefont {Zyczkowski}}]{Kukulski_2021}%
  \BibitemOpen
  \bibfield  {author} {\bibinfo {author} {\bibfnamefont {R.}~\bibnamefont
  {Kukulski}}, \bibinfo {author} {\bibfnamefont {I.}~\bibnamefont {Nechita}},
  \bibinfo {author} {\bibfnamefont {L.}~\bibnamefont {Pawela}}, \bibinfo
  {author} {\bibfnamefont {Z.}~\bibnamefont {Puchala}},\ and\ \bibinfo {author}
  {\bibfnamefont {K.}~\bibnamefont {Zyczkowski}},\ }\bibfield  {title}
  {\bibinfo {title} {Generating random quantum channels},\ }\href
  {http://dx.doi.org/10.1063/5.0038838} {\bibfield  {journal} {\bibinfo
  {journal} {Journal of Mathematical Physics}\ }\textbf {\bibinfo {volume}
  {62}} (\bibinfo {year} {2021})}\BibitemShut {NoStop}%
\bibitem [{\citenamefont {Inc.}(2022)}]{MATLAB}%
  \BibitemOpen
  \bibfield  {author} {\bibinfo {author} {\bibfnamefont {T.~M.}\ \bibnamefont
  {Inc.}},\ }\href {https://www.mathworks.com} {\bibinfo {title} {Matlab
  version: 9.13.0 (r2022b)}} (\bibinfo {year} {2022})\BibitemShut {NoStop}%
\bibitem [{\citenamefont {Grant}\ and\ \citenamefont {Boyd}(2014)}]{cvx}%
  \BibitemOpen
  \bibfield  {author} {\bibinfo {author} {\bibfnamefont {M.}~\bibnamefont
  {Grant}}\ and\ \bibinfo {author} {\bibfnamefont {S.}~\bibnamefont {Boyd}},\
  }\href@noop {} {\bibinfo {title} {{CVX}: Matlab software for disciplined
  convex programming, version 2.1}},\ \bibinfo {howpublished}
  {\url{http://cvxr.com/cvx}} (\bibinfo {year} {2014})\BibitemShut {NoStop}%
\bibitem [{\citenamefont {Johnston}(2016)}]{qetlab}%
  \BibitemOpen
  \bibfield  {author} {\bibinfo {author} {\bibfnamefont {N.}~\bibnamefont
  {Johnston}},\ }\href {https://doi.org/10.5281/zenodo.44637} {\bibinfo {title}
  {{QETLAB}: A {MATLAB} toolbox for quantum entanglement, version 0.9}},\
  \bibinfo {howpublished} {\url{https://qetlab.com}} (\bibinfo {year}
  {2016})\BibitemShut {NoStop}%
\bibitem [{\citenamefont {Wootters}(1987)}]{WOOTTERS19871}%
  \BibitemOpen
  \bibfield  {author} {\bibinfo {author} {\bibfnamefont {W.~K.}\ \bibnamefont
  {Wootters}},\ }\bibfield  {title} {\bibinfo {title} {A wigner-function
  formulation of finite-state quantum mechanics},\ }\href
  {https://doi.org/https://doi.org/10.1016/0003-4916(87)90176-X} {\bibfield
  {journal} {\bibinfo  {journal} {Annals of Physics}\ }\textbf {\bibinfo
  {volume} {176}},\ \bibinfo {pages} {1} (\bibinfo {year} {1987})}\BibitemShut
  {NoStop}%
\bibitem [{\citenamefont {Gross}(2006{\natexlab{a}})}]{Gross_2006a}%
  \BibitemOpen
  \bibfield  {author} {\bibinfo {author} {\bibfnamefont {D.}~\bibnamefont
  {Gross}},\ }\bibfield  {title} {\bibinfo {title} {Hudson's theorem for
  finite-dimensional quantum systems},\ }\bibfield  {journal} {\bibinfo
  {journal} {Journal of Mathematical Physics}\ }\textbf {\bibinfo {volume}
  {47}},\ \href {https://doi.org/10.1063/1.2393152} {10.1063/1.2393152}
  (\bibinfo {year} {2006}{\natexlab{a}})\BibitemShut {NoStop}%
\bibitem [{\citenamefont {Gross}(2006{\natexlab{b}})}]{Gross_2006b}%
  \BibitemOpen
  \bibfield  {author} {\bibinfo {author} {\bibfnamefont {D.}~\bibnamefont
  {Gross}},\ }\bibfield  {title} {\bibinfo {title} {Non-negative wigner
  functions in prime dimensions},\ }\href
  {https://doi.org/10.1007/s00340-006-2510-9} {\bibfield  {journal} {\bibinfo
  {journal} {Applied Physics B}\ }\textbf {\bibinfo {volume} {86}},\ \bibinfo
  {pages} {367} (\bibinfo {year} {2006}{\natexlab{b}})}\BibitemShut {NoStop}%
\bibitem [{\citenamefont {Kristj{\'{a}}nsson}\ \emph
  {et~al.}(2020)\citenamefont {Kristj{\'{a}}nsson}, \citenamefont {Chiribella},
  \citenamefont {Salek}, \citenamefont {Ebler},\ and\ \citenamefont
  {Wilson}}]{Kristj_nsson_2020}%
  \BibitemOpen
  \bibfield  {author} {\bibinfo {author} {\bibfnamefont {H.}~\bibnamefont
  {Kristj{\'{a}}nsson}}, \bibinfo {author} {\bibfnamefont {G.}~\bibnamefont
  {Chiribella}}, \bibinfo {author} {\bibfnamefont {S.}~\bibnamefont {Salek}},
  \bibinfo {author} {\bibfnamefont {D.}~\bibnamefont {Ebler}},\ and\ \bibinfo
  {author} {\bibfnamefont {M.}~\bibnamefont {Wilson}},\ }\bibfield  {title}
  {\bibinfo {title} {Resource theories of communication},\ }\href
  {https://doi.org/10.1088/1367-2630/ab8ef7} {\bibfield  {journal} {\bibinfo
  {journal} {New Journal of Physics}\ }\textbf {\bibinfo {volume} {22}},\
  \bibinfo {pages} {073014} (\bibinfo {year} {2020})}\BibitemShut {NoStop}%
\bibitem [{\citenamefont {Mari}\ and\ \citenamefont {Eisert}(2012)}]{Mari2012}%
  \BibitemOpen
  \bibfield  {author} {\bibinfo {author} {\bibfnamefont {A.}~\bibnamefont
  {Mari}}\ and\ \bibinfo {author} {\bibfnamefont {J.}~\bibnamefont {Eisert}},\
  }\bibfield  {title} {\bibinfo {title} {{Positive wigner functions render
  classical simulation of quantum computation efficient}},\ }\href
  {https://doi.org/10.1103/PhysRevLett.109.230503} {\bibfield  {journal}
  {\bibinfo  {journal} {Physical Review Letters}\ }\textbf {\bibinfo {volume}
  {109}},\ \bibinfo {pages} {1} (\bibinfo {year} {2012})},\ \Eprint
  {https://arxiv.org/abs/1208.3660} {arXiv:1208.3660} \BibitemShut {NoStop}%
\bibitem [{\citenamefont {Veitch}\ \emph {et~al.}(2012)\citenamefont {Veitch},
  \citenamefont {Ferrie}, \citenamefont {Gross},\ and\ \citenamefont
  {Emerson}}]{Veitch2012}%
  \BibitemOpen
  \bibfield  {author} {\bibinfo {author} {\bibfnamefont {V.}~\bibnamefont
  {Veitch}}, \bibinfo {author} {\bibfnamefont {C.}~\bibnamefont {Ferrie}},
  \bibinfo {author} {\bibfnamefont {D.}~\bibnamefont {Gross}},\ and\ \bibinfo
  {author} {\bibfnamefont {J.}~\bibnamefont {Emerson}},\ }\bibfield  {title}
  {\bibinfo {title} {{Negative quasi-probability as a resource for quantum
  computation}},\ }\href {https://doi.org/10.1088/1367-2630/14/11/113011}
  {\bibfield  {journal} {\bibinfo  {journal} {New Journal of Physics}\ }\textbf
  {\bibinfo {volume} {14}},\ \bibinfo {pages} {1} (\bibinfo {year} {2012})},\
  \Eprint {https://arxiv.org/abs/1201.1256} {arXiv:1201.1256} \BibitemShut
  {NoStop}%
\bibitem [{\citenamefont {Wang}\ and\ \citenamefont {Wilde}(2019)}]{WW19}%
  \BibitemOpen
  \bibfield  {author} {\bibinfo {author} {\bibfnamefont {X.}~\bibnamefont
  {Wang}}\ and\ \bibinfo {author} {\bibfnamefont {M.~M.}\ \bibnamefont
  {Wilde}},\ }\bibfield  {title} {\bibinfo {title} {{Resource theory of
  asymmetric distinguishability}},\ }\href
  {https://doi.org/10.1103/PhysRevResearch.1.033170} {\bibfield  {journal}
  {\bibinfo  {journal} {Physical Review Research}\ }\textbf {\bibinfo {volume}
  {1}},\ \bibinfo {pages} {033170} (\bibinfo {year} {2019})},\ \Eprint
  {https://arxiv.org/abs/1905.11629} {arXiv:1905.11629} \BibitemShut {NoStop}%
\bibitem [{\citenamefont {Gour}(2024)}]{gour2024resources}%
  \BibitemOpen
  \bibfield  {author} {\bibinfo {author} {\bibfnamefont {G.}~\bibnamefont
  {Gour}},\ }\href@noop {} {\bibinfo {title} {Resources of the quantum world}}
  (\bibinfo {year} {2024}),\ \Eprint {https://arxiv.org/abs/2402.05474}
  {arXiv:2402.05474 [quant-ph]} \BibitemShut {NoStop}%
\end{thebibliography}
\end{document}